\documentclass[11pt]{article}
\pdfoutput=1

\usepackage{euscript}
\usepackage{amsfonts}
\usepackage{amsbsy}
\usepackage{epsfig}
\usepackage{amsthm}
\usepackage{amscd}
\usepackage{amstext}
\usepackage{verbatim}
\usepackage{cancel}
\usepackage{capt-of}
\usepackage{empheq}

\usepackage[nosort]{cite}
\usepackage{bm}
\usepackage{authblk}
\usepackage{esint}
\usepackage[T1]{fontenc}
\usepackage{mathdots}
\usepackage[centertableaux]{ytableau}

\usepackage[utf8]{inputenc}
\usepackage{graphicx}
\usepackage{physics}
\usepackage{mathtools}
\usepackage{bbold}

\usepackage{setspace}
\usepackage{colortbl}
\usepackage{xcolor}

\usepackage{amsmath}
\makeatletter
\renewcommand*\env@matrix[1][\arraystretch]{%
  \edef\arraystretch{#1}%
  \hskip -\arraycolsep
  \let\@ifnextchar\new@ifnextchar
  \array{*\c@MaxMatrixCols c}}
\makeatother
\usepackage{amssymb}
\usepackage{mathrsfs}
\usepackage{mathtools}
\usepackage{physics}
\usepackage{booktabs}
\usepackage{bbm}
\usepackage{float}

\usepackage{pgfplots}
\pgfplotsset{compat=1.15}
\usepackage{tikz}
\usetikzlibrary{external}
\usetikzlibrary{arrows}

\usepackage{subcaption}
\usepackage{graphicx}
\usepackage{mathtools}
\usepackage[many]{tcolorbox}
\tcbset{shield externalize}
\usepackage{xcolor, colortbl}
\usepackage[a4paper]{geometry}
\usepackage[hidelinks,linktocpage]{hyperref}
\hypersetup{
    colorlinks=true,
    linkcolor=blue,
    citecolor=blue,
    }

\usepackage{mathrsfs}

\def\ben{\begin{equation}}
\def\een{\end{equation}}
\def\half{{\textstyle{\frac{1}{2}}}}
\let\a=\alpha    
   
\let\l=\lambda   \let\x=\xi

\let\w=\omega   
   
\let\C=\Chi

\let\pa=\partial
\def\be{\begin{equation}}
\def\ee{\end{equation}}
\def\beq{\begin{equation}}
\def\eeq{\end{equation}}
\def\ba{\begin{array}}
\def\ea{\end{array}}

\def\dalemb#1#2{{\vbox{\hrule height .#2pt
       \hbox{\vrule width.#2pt height#1pt \kern#1pt
               \vrule width.#2pt}
       \hrule height.#2pt}}}

\newcommand{\bea}{\begin{eqnarray}}
\newcommand{\eea}{\end{eqnarray}}

\def\vep{{\varepsilon}}

\DeclareMathOperator{\ord}{ord}
\makeatletter
\newcommand*\bigcdot{\mathpalette\bigcdot@{.5}}
\newcommand*\bigcdot@[2]{\mathbin{\vcenter{\hbox{\scalebox{#2}{$\m@th#1\bullet$}}}}}
\makeatother

\def\R{{{\mathbb R}}}
\def\P{{{\mathbb P}}}
\def\C{{{\Bbb C}}}
\def\Q{{{\Bbb Q}}}
\def\H{{{\Bbb H}}}
\def\O{{{\Bbb O}}}

\def\Z{{{\mathbb Z}}}

\def\N{{{\mathbb N}}}

\def\ocal{{\mathcal{O}}}

\def\cp{{\mathfrak{p}}}

\title{Wheeler-DeWitt wavefunctions for 5d BKL dynamics, \\
automorphic $L$-functions and complex primon gases
}

\author{Marine De Clerck, Sean~A.~Hartnoll and Ming~Yang}
\affil{\it Department of Applied Mathematics and Theoretical Physics, \\
\it University of Cambridge, Cambridge CB3 0WA, UK
}
\date{}

\begin{document}

\maketitle

\begin{abstract}

The near-singularity BKL dynamics of five dimensional gravity and supergravity (and also an extended four-dimensional supergravity) is known to be given by the billiard problem of a particle within a fundamental domain of the Bianchi groups $PSL(2,\ocal) \subset PSL(2,\C)$, acting on $\H_3$. Here $\ocal$ are the Gaussian or Eisenstein integers, which define a square or triangular lattice in $\C$. The Wheeler-DeWitt wavefunctions near the singularity are, correspondingly, automorphic Maa{\ss} forms of $PSL(2,\ocal)$. We show how these wavefunctions are associated to certain $L$-functions evaluated along their critical axis. Each of these $L$-functions admits an Euler product representation over the complex primes ${\mathcal{P}}_\ocal \subset \ocal$. From this fact we write the $L$-function as the trace over an auxiliary Hilbert space of charged harmonic oscillators, labeled by the complex primes ${\mathcal{P}}_\ocal$. In this way we have constructed a `dual' primon gas partition function for the wavefunction of the universe close to a five dimensional cosmological singularity.

\end{abstract}

\newpage

\tableofcontents

\section{Introduction}

\subsection{Primon gases for four-dimensional gravity}
\label{sec:4d}

It was argued by Belinsky-Khalatnikov-Lifshitz (BKL) that close to a spacelike singularity the equations of general relativity simplify dramatically \cite{Belinsky:1970ew}. Spatial points decouple from one another and evolve separately following a conservative chaotic dynamics, later called the cosmological billiard \cite{Damour:2002et}. It is remarkable that, for four-dimensional gravity, the billiard domain is half of the fundamental domain of $PSL(2,\Z) \subset PSL(2,\R)$ acting on the hyperbolic plane $\H_2$ \cite{belinski_henneaux_2017}. This fact connects the gravitational evolution to deep and well-studied mathematics.

The quantisation of the billiard problem gives a semiclassical Wheeler-DeWitt (WDW) quantisation of gravity close to a singularity \cite{DeWitt:1967yk}. There is no minisuperspace approximation in the BKL limit, as each point in space has decoupled and can be treated separately. Throughout, we will be considering the quantum dynamics of the metric at a fixed spatial point. The quantum eigenmodes $\psi_\text{WDW}(x,y)$ of the billiard problem are precisely the odd automorphic forms of $PSL(2,\Z)$, known as Maa{\ss} waveforms. These are central objects in the field of arithmetic quantum chaos \cite{Bogomolny:1992cj, sarnak1993arithmetic, em/1048610117}. Early work connecting quantum cosmology to arithmetic chaos, mostly in the context of minisuperspace models, is reviewed in \cite{GRAHAM19951103}.

We have recently established that every automorphic Wheeler-DeWitt eigenform has a corresponding primon gas partition function, given by an $L$-function \cite{Hartnoll:2025hly}:
\be\label{eq:map1}
\psi_\text{WDW}(x,y) \quad \leftrightarrow \quad L(s) = \text{Tr}_{\mathcal{H}}\, e^{- s H + i\theta \cdot  Q}  \,.
\ee
Here the trace is over an auxiliary Hilbert space of non-interacting, charged oscillators labeled by prime numbers:
\be\label{eq:Hp}
{\mathcal H} = \bigotimes_{p \in \P} {\mathcal H}_p \,.
\ee
The operators $H$ and $\theta \cdot Q$ act diagonally with respect to the decomposition (\ref{eq:Hp}). The original primon gas established an analogous Hilbert space interpretation of the Riemann zeta function \cite{mackey1978unitary, julia1990statistical}. 
The key mathematical input is the Euler product formula that transforms the partition function sum over integers into a product over primes. The connection in (\ref{eq:map1}) is that the chemical potentials $\{\theta_p\}$ determine the Fourier coefficients of the corresponding Maa{\ss} waveform.

A direct relationship to the $L$-function arises by taking a Mellin transform of the eigenform along the radial direction. Setting the $\H_2$ coordinates $x+i y = \rho e^{i \theta}$ \cite{Epstein1985, Hafner1987}:
\be\label{eq:map2}
L(\half + i t) \quad \propto \quad \int_0^\infty \frac{d\rho}{\rho} \rho^{it} \psi_\text{WDW}(\rho \cos\theta, \rho\sin\theta)  \,.
\ee
It is explained in \cite{Hartnoll:2025hly} that this Mellin transform is equivalent to expressing the wavefunction in a basis of hyperbolic dilatation eigenstates. Whereas the primon gas partition function in (\ref{eq:map1}) involves $L(s)$ along the real axis, the $L$-function in (\ref{eq:map2}) is evaluated along the critical axis. According to the (generalised) Riemann hypothesis, this axis contains the nontrivial zeros $\{t_n\}$ of the $L$-function. The $L$-function can be written as a Hadamard product over these zeros. The nontrivial zeros obey Wigner-Dyson statistics \cite{RS, rudnick}, suggestive of an interpretation as the spectrum of an auxiliary complicated Hamiltonian. This Hamiltonian is distinct from the primon gas Hamiltonian and will not play a role in our discussion.

In each of (\ref{eq:map1}) and (\ref{eq:map2}) above the data in the wavefunction of the universe is packaged into a collection of numbers, $\{\theta_p\}$ or $\{t_n\}$. This data, we have noted, is associated to auxiliary Hilbert spaces. The association of a partition function to a quantum state, in particular, has had powerful consequences for gapped states with topological order \cite{RevModPhys.89.025005} and is also at the heart of the dS/CFT correspondence \cite{Strominger:2001pn, Maldacena:2002vr}. Our hope is that primon gases are the foundation of a dual formulation that could similarly illuminate the fate of spacetime at the singularity.

\subsection{Cosmological billiards above four dimensions}

The classical description of gravity will ultimately break down as the singularity is approached. To capture this emergence of microscopic physics, the dual primon gas partition functions (and/or Hamiltonians for the nontrivial zeros) will presumably need to be enhanced to incorporate stringy degrees of freedom and quantum gravitational effects. To understand how this works, it is natural to consider the cosmological billiards of a gravitational theory with a known UV completion.
This approach has been taken in a series of works that have grappled with the quantum cosmological billiards of eleven dimensional supergravity \cite{Kleinschmidt:2009hv, Kleinschmidt:2009cv, Kleinschmidt:2022qwl}.\footnote{Within pure gravity, BKL chaos persists up to ten spacetime dimensions \cite{Elskens:1987rk, Elskens:1987gj}.}
Again remarkably, in this case the billiard dynamics unfolds within the fundamental domain in $\H_9$ of a modular group over the integral octonians, called octavians.

Our eventual objective is to develop a primon gas description of the Maa{\ss} waveforms for eleven dimensional supergravity. The theory of modular invariance over the octonians $\O$ remains to be fully developed. However, there is a natural path from the $\H_2$ cosmological billiard of four-dimensional gravity to the $\H_9$ cosmological billiard of eleven dimensional supergravity. It has been shown that there is a close connection between the integers of normed division algebras ($\R,\C,\H,\O$) and generalised modular groups \cite{Feingold:2008ih, Kleinschmidt:2010bk}. The corresponding fundamental domains are within $\H_2, \H_3,\H_5$ and $\H_9$ and all arise as the cosmological billiard of simple gravitational theories \cite{Cremmer:1999du,Damour:2002fz,Feingold:2008ih}. In this paper we take the first step along this path and derive the dual primon gas description of $\H_3$ cosmological billiards. The billiard will be the fundamental domain of certain modular subgroups of $PSL(2,\C)$, arising from simple gravitational theories.

We will consider three gravitational theories. These are (i) pure gravity in five dimensions, (ii) Einstein-Maxwell theory in five dimensions and (iii) Einstein-Maxwell-axion-dilaton (EMAD) theory in four dimensions.
The latter two cases also arise as supergravity theories, as we describe below.
In \S\ref{sec:five} we review the cosmological billiard for these theories, obtained in \cite{Damour:2002et, Damour:2002fz}. Using coordinates $(x_1,x_2,y)$ for $\H_3$, the billiard domains involve triangular subregions of the complex plane $x_1 + i x_2$ that are associated to the Eisenstein and Gaussian integers, $\Z[\omega]$ and $\Z[i]$. These are sets of complex integers, corresponding to the simplest triangular and square lattices in $\C$. A crucial fact for us will be that these integers admit prime factorisation over a set of complex prime numbers. We will denote the integers by $\ocal$ and the corresponding primes by ${\mathcal P}_\ocal$.

With the billiard domains in hand, in \S\ref{sec:auto} we review the construction of the WDW wavefunctions $\psi_\text{WDW}(x_1,x_2,y)$ for the cosmological billiards (e.g.~\cite{steil1996eigenvalues,Aurich:2004ik}). Similarly to the case of four dimensional gravity, these are odd Maa{\ss} waveforms, now invariant under $PSL(2,\ocal)$.

In \S\ref{sec:Lfun} we express the wavefunctions in terms of $L$-functions, generalising (\ref{eq:map2}) to our higher dimensional setting. Because $SL(2,\C)$ includes commuting rotation and dilatation generators, each wavefunction leads to family of $L$-functions labeled by an angular momentum eigenvalue $n$ in addition to a dilatation eigenvalue $t$. Introducing polar coordinates $(\rho,\vartheta,\varphi)$ on $\H_3$,
\be\label{eq:map3}
L_n(1 + i t) \quad \propto \quad \int_0^\infty \frac{d\rho}{\rho} \rho^{it} \int_0^{2\pi} d\varphi \, e^{i n \varphi}\, \psi_\text{WDW}(\rho\sin\vartheta\cos\varphi, \rho\sin\vartheta\sin\varphi,\rho \cos\vartheta)  \,.
\ee
The $L$-functions are again evaluated along their critical axis, which is expected to contain the nontrivial zeroes of the function.

In \S\ref{sec:primon} we review the Euler product formula for our $PSL(2,\ocal)$ automorphic waveforms, via Hecke relations for the Fourier coefficients (e.g.~\cite{steil1996eigenvalues}). This allows us to associate the WDW wavefunction to a primon gas partition function
\be\label{eq:con2}
\psi_\text{WDW}(x_1,x_2,y) \quad \leftrightarrow \quad L_n(s) = \text{tr}_\mathcal{H} \exp{- s H + i \theta \cdot Q - i n L}\,.
\ee
This relationship is the result that our paper build up to. The partition function in (\ref{eq:con2}) generalises the primon gas (\ref{eq:map1}) to a gas of oscillators labelled by complex primes, so that the Hilbert space:
\be\label{eq:Hp2}
{\mathcal H} = \bigotimes_{\cp \in {\mathcal P}_\ocal} {\mathcal H}_\cp \,.
\ee
As in the previous \S\ref{sec:4d}, the operators $H, \theta \cdot Q$ and $L$ are diagonal with respect to the decomposition (\ref{eq:Hp2}). The partition function in (\ref{eq:con2}) is therefore a product of partitions functions $z_\cp$ for the individual primes.

In the remainder of \S\ref{sec:primon} we make a few observations about the complex primon gas. We show that the oscillator partition function for a single complex prime can be written in the suggestive form $z_\cp = \tr \left(\left(1/\cp\right)^{A^\cp_{s+n}} \left(1/\bar{\cp}\right)^{A^\cp_{s-n}}\right)$, for operators $A^\cp_{s\pm n}$ that have some resemblance to Virasoro charges $\{L_0,\overline{L}_0\}$. We also perform numerical computations of the prime Fourier coefficients for a fixed wavefunction and verify that the associated `Hecke eigenvalues' follow a Sato-Tate semicircle distribution. Finally, we use the known Kesten-McKay distribution of prime Fourier coefficients among different automorphic $L$-functions to average the primon gases over different WDW energy levels. As in \cite{Hartnoll:2025hly}, the averaged primon gas has a fermionc character and the averaged oscillators acquire a degeneracy that grows with $|\cp|$.
We end with a few comments in \S\ref{sec:final}.

\section{Cosmological billiards in $\H_3$}
\label{sec:five}

In this section we will review the cosmological billiards of three gravitational theories, quoting results from \cite{Damour:2002et, Feingold:2008ih, Damour:2002fz}. These theories have in common that the billiard domain is within hyperbolic upper half-space,
\begin{equation}\label{eq:H31}
    \mathbb{H}_3 \equiv \left\{ (x_1, x_2, y) \in \mathbb{R}^3 \ \middle| \ y > 0 \right\} \,.
\end{equation}
This space is 
equipped with the metric
\begin{equation}\label{eq:H32}
    ds^2 = \frac{dx_1^2 + dx_2^2 + dy^2}{y^2} \,.
\end{equation}

\subsection{$5d$ gravity and Eisenstein integers}
\label{sec:E}

Pure five-dimensional gravity has the action
\begin{equation}
    S =  \int \mathrm{d}^5x \sqrt{-g}R \,.
\end{equation}
The corresponding cosmological billiard is obtained from the Hamiltonian constraint. To write down this constraint, the four-metric on a spatial slice can be written using an Iwasawa decomposition as \cite{Damour:2002et}
\be
ds^2_4 = \sum_{i=1}^4 e^{- 2 \beta_i(x)} \theta_i(x)^2 \,.
\ee
Here the $\beta_i(x)$ are functions on the spatial slice and the $\theta_i(x) = \sum_j \theta_{ij}(x) dx_j$ are one-forms such that $\theta_{ij}$ is an upper-diagonal matrix with unit entries along the diagonal. In particular, this means that the local volume of the spatial slice is $e^{-\sum_i \beta_i(x)}$. Towards the singularity the volume goes to zero. To leading order in this limit it is
found that the $\theta_i(x)$ freeze while the $\beta_i(x)$ undergo free motion that is bounded by `walls' \cite{Damour:2002et}. For any given fixed spatial point $x$, let $\pi_i$ be the momentum conjugate to $\beta_i$. The free motion is determined by
the Hamiltonian constraint, which takes the form (for each spatial point independently)
\be\label{eq:pipi}
\sum_{i=1}^4 \pi_i^2 - \frac{1}{3}\left(\sum_{i=1}^4 \pi_i \right)^2 = 0 \,.
\ee
At the walls, the motion undergoes an instantaneous bounce.

\begin{figure}[h]
\centering
\includegraphics[width=0.3\textwidth]{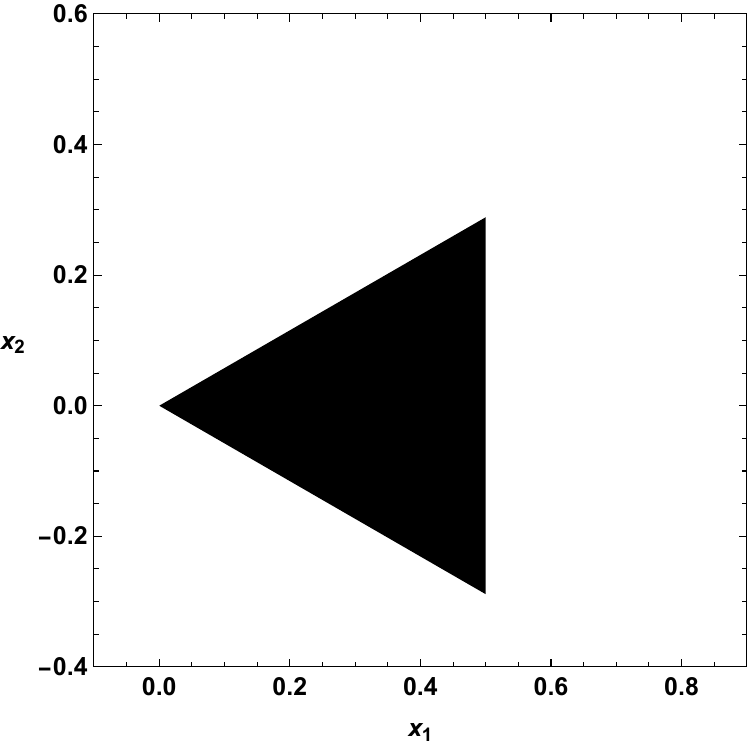}\hspace{0.5cm}
\includegraphics[width=0.3\textwidth]{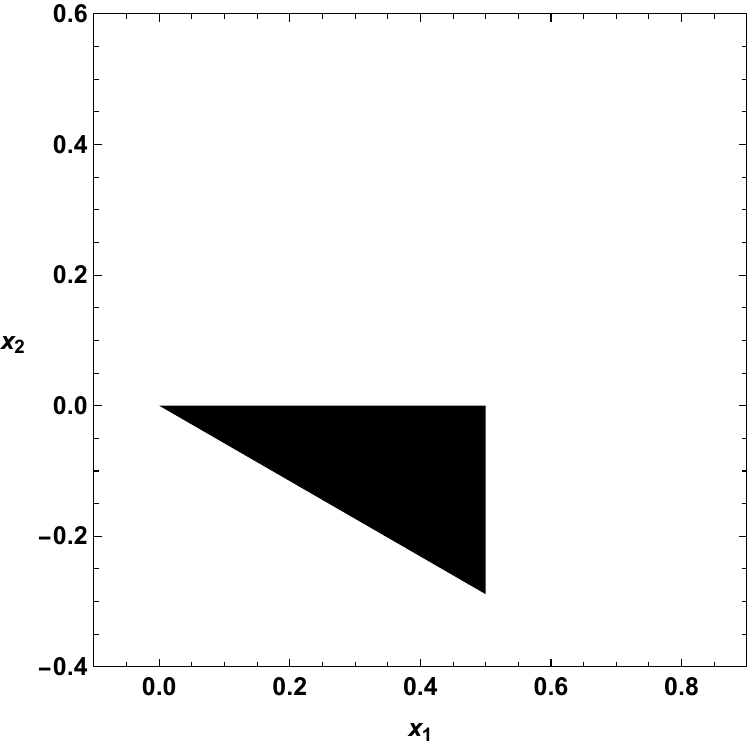}
\hspace{0.5cm}
\includegraphics[width=0.3\textwidth]{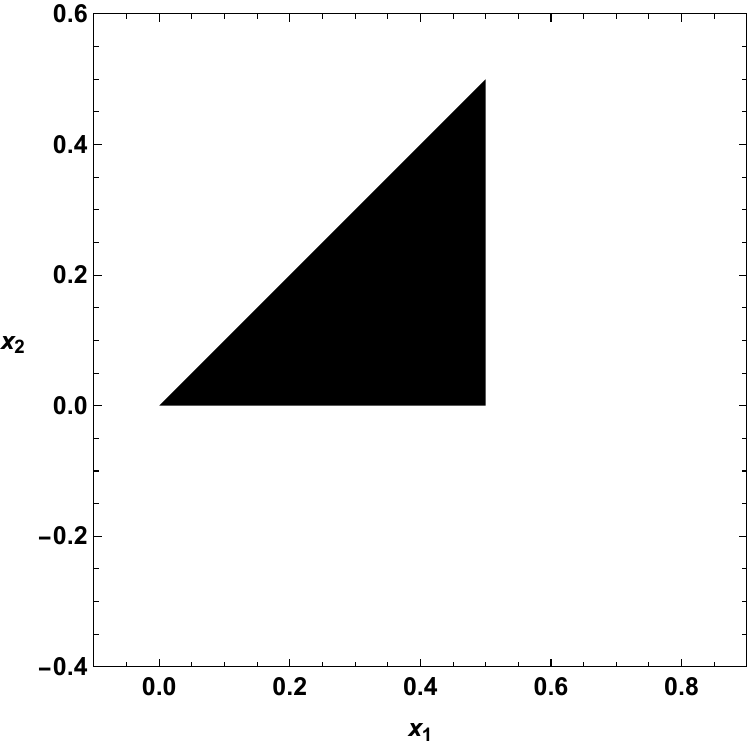}
\caption{Triangular domains in the $(x_1,x_2)$ plane for the cosmological billiards of the three theories discussed. From left to right: 5d gravity (equilateral), 5d Einstein-Maxwell (hemi-equilateral) and 4d Einstein-Maxwell-axion-dilaton (isosceles right-angled). The first two domains relate to Eisenstein integers while the final one relates to Gaussian integers.
\label{fig:triangle}}
\end{figure}

In pure five-dimensional gravity there are `curvature' and `symmetry' walls. Taken together, the set of dominant walls is
\begin{equation} \label{eq: walls gravity 5d}
    \beta_1 \leq \beta_2 \leq 
    \beta_3 \leq \beta_4 , \qquad 0 \leq 2\beta_1+\beta_2 \,.  
\end{equation}
These restrict the scale factors to a region in superspace. This region is made more transparent by the following change of variables,
\begin{align}\label{eq: beta to UHP 1}
    \beta_1 & = \frac{-2x_2}{\sqrt{6}y}e^{\tau} \,,
    && \beta_2 = \frac{\sqrt{3}x_1+x_2}{\sqrt{6}y}e^{\tau} \,, \\
    \beta_3 & = \frac{\sqrt{3}(1-x_1)+x_2}{\sqrt{6}y}e^{\tau}
    \,,
    && \beta_4 = \frac{-\sqrt{3}x_1+x_2+\sqrt{3}(x_1^2+x_2^2+y^2)}{\sqrt{6}y}e^{\tau} \,. \nonumber
\end{align}
This transformation is similar to going to Milne coordinates, and has two consequences. Firstly, the Hamiltonian constraint (\ref{eq:pipi}) becomes
\begin{equation}\label{eq:Hamiltonian}
    \pi_\tau^2 = y^2(\pi_{x_1}^2+\pi_{x_2}^2+ \pi_y^2) \,.
\end{equation}
Secondly, the walls \eqref{eq: walls gravity 5d} become
\begin{equation}\label{eq:bdy:A2}
    x_1^2+x_2^2+y^2 \geq 1, \qquad
    x_1 - \sqrt{3}x_2 \geq 0, \qquad
    x_1 \leq \frac{1}{2}, \qquad
    x_1+\sqrt{3}x_2 \geq 0\,.
\end{equation}
The expression (\ref{eq:Hamiltonian}) says that evolution in $\tau$ is given by free motion on three dimensional hyperbolic space, $\H_3$, defined in (\ref{eq:H31}) and (\ref{eq:H32}). The walls in (\ref{eq:bdy:A2}) are all independent of $\tau$. The final three walls in (\ref{eq:bdy:A2}) define an equilateral triangle in the $(x_1,x_2)$ plane, illustrated in Fig.~\ref{fig:triangle}. Superspace evolution must be within this triangle. We will return to the first wall in (\ref{eq:bdy:A2}) later.

It will be crucial later that the equilateral triangle is closely related to the Eisenstein integers
\be\label{eq:eisenstein}
\Z[\omega] \equiv \Z + \omega \Z \,, \qquad \omega = e^{2 \pi i/3}\,.
\ee
These integers form a triangular lattice in $\C$, related to the root system of the Lie algebra $A_2$. In \S\ref{sec:eisdom} we will see that the interior of the equilateral triangle in Fig.~\ref{fig:triangle} is the fundamental domain of the symmetry group of this lattice acting on $\C$. This symmetry group is generated by translations by Eisenstein integers and rotations by $\pm \omega$. Upon incorporating the $y$ direction, the full billiard (\ref{eq:bdy:A2}) will be seen to be the fundamental domain of a modular group over $\Z[\omega]$, now acting on $\H_3$.

\subsection{$5d$ Einstein-Maxwell and Eisenstein integers}
\label{sec:EM}

A Maxwell field can be added to the theory we have just considered,
\begin{equation}\label{eq:EMth}
    S =  \int \mathrm{d}^5x \sqrt{-g}\left( R - \frac{1}{4} F_{\mu\nu}F^{\mu\nu}\right).
\end{equation}
A five-dimensional Chern-Simons term can further be added to the Einstein-Maxwell theory (\ref{eq:EMth}), giving the bosonic sector of minimal $5d$ supergravity. The Chern-Simons term does not alter the dominant walls below and hence our considerations also apply to the supergravity theory \cite{Damour:2002fz}.

The Maxwell field leads to additional `electric' walls, while itself freezing towards the singularity. Thus, the cosmological billiard dynamics involves only the four scale factors considered above. The dominant walls are now found to be
\begin{equation}\label{eq:EMwalls}
0 \leq \beta_1 \leq \beta_2 \leq 
    \beta_3 \leq \beta_4 \,.
\end{equation}

As previously, the change of variables (\ref{eq: beta to UHP 1}) maps the Hamiltonian constraint to the hyperbolic form (\ref{eq:Hamiltonian}). The walls (\ref{eq:EMwalls}) become
\begin{equation}\label{eq:bdy:G2}
    x_1^2+x_2^2+y^2 \geq 1, \qquad
    x_1  \leq \frac{1}{2}, \qquad
    x_1+\sqrt{3}x_2 \geq 0, \qquad
    x_2  \leq 0.
\end{equation}
The allowed triangular domain in the $(x_1,x_2)$ plane is shown in Fig.~\ref{fig:triangle}. It is a hemi-equilateral triangle, half of the equilateral domain we found previously for pure five-dimensional gravity. This smaller triangle is also closely related to the Eisenstein integers, as we will explain in \S\ref{sec:eisdom}, and to the root system of the Lie algebra $G_2$. Upon extending to include the $y$ direction, the full billiard (\ref{eq:bdy:G2}) will again be the fundamental domain of a modular group over $\Z[\omega]$, acting on $\H_3$.

\subsection{$4d$ Einstein-Maxwell-axion-dilaton and Gaussian integers}
\label{sec:EMAD}

Consider now four-dimensional gravity coupled to a Maxwell field, an axion and a dilaton \cite{Damour:2002fz}
\begin{equation}\label{eq:emd}
     S =  \int \mathrm{d}^4x \sqrt{-g} \left( R - \frac{e^{\sqrt{2}\phi}}{4} F_{\mu\nu}F^{\mu\nu} - \frac{e^{2\sqrt{2}\phi}}{2} \pa^\mu \chi \pa_\mu \chi -\pa^\mu \phi \pa_\mu \phi  \right).
\end{equation}
This Lagrangian is equivalent, in the billiard regime where the axion $\chi$ becomes constant, to pure four dimensional $\mathcal{N}=4$ supergravity truncated to a single graviphoton \cite{Gibbons:1982ih}, after performing an $SL(2,\R)$ transformation \cite{kallosh1994supersymmetry}.

In (\ref{eq:emd}), the axion and Maxwell field freeze near the singularity, while the dilaton undergoes free motion together with the three scale factors. The Hamiltonian constraint is now \cite{Damour:2002et}
\be\label{eq:pipi2}
\sum_{i=1}^3 \pi_i^2 - \frac{1}{2}\left(\sum_{i=1}^3 \pi_i \right)^2 + \pi_\phi^2 = 0 \,.
\ee
Here $\pi_\phi$ is the momentum conjugate to the dilaton.
The dominant restrictions on the motion come from the symmetry walls, the `electric' wall for the axion and the `magnetic' wall for the Maxwell field, such that \cite{Damour:2002fz}
\be\label{eq:axiowalls}
0 \leq \frac{\phi}{\sqrt{2}} \leq \beta_1 \leq \beta_2 \leq \beta_3 \,.
\ee

The dominant walls in (\ref{eq:axiowalls}) are similar to (\ref{eq:EMwalls}) previously, but the Hamiltonian constraint (\ref{eq:pipi2}) is not the same as (\ref{eq:pipi}). Therefore, a different change of variables is needed to map the problem onto motion in $\H_3$. We may set
\begin{align}
    \beta_1 = \frac{x_1}{\sqrt{2}y}e^{\tau} \,, \quad
    \beta_2 = \frac{1-x_1}{\sqrt{2}y}e^{\tau} \,, \quad
    \beta_3 = \frac{-x_1+x^2_1+x^2_2+y^2}{\sqrt{2}y}e^{\tau} \,, \quad \phi = \frac{x_2}{y}e^{\tau} \,.
\end{align}
This maps the Hamiltonian constraint (\ref{eq:pipi2}) into the hyperbolic form (\ref{eq:Hamiltonian}). The walls become
\begin{equation}\label{eq:bdy:C2}
    x_1^2+x_2^2+y^2 \geq 1, \qquad
    x_1  \leq \frac{1}{2}, \qquad
    x_1 - x_2 \geq 0, \qquad
    x_2  \geq 0.
\end{equation}
The allowed domain in the $(x_1,x_2)$ plane is
an isosceles right-angled triangle, shown in Fig.~\ref{fig:triangle}.

The right-angled triangular domain in Fig.~\ref{fig:triangle} is now closely related to the Gaussian integers
\be\label{eq:gaussian}
\Z[i] \equiv \Z + i \Z \,.
\ee
These integers form a square lattice in $\C$, related to the root system of the Lie algebra $C_2$. The symmetry group of this lattice is generated by translations by Gaussian integers and rotations by $i$. In \S\ref{sec:gaussdom} below we explain that the interior of the right-angled triangle in Fig.~\ref{fig:triangle} is the fundamental domain of this symmetry group, enhanced by a further reflection. Upon incorporating the $y$ direction, the full billiard (\ref{eq:bdy:C2}) will be seen to be the fundamental domain of a modular group over $\Z[i]$, acting on $\H_3$.

\section{Automorphic forms on $\H_3$}
\label{sec:auto}

\subsection{Wave equation and mode expansions}

Wheeler-DeWitt quantisation near the singularity means a canonical quantisation of the Hamiltonian constraint (\ref{eq:Hamiltonian}). This suffers from an ordering ambiguity and also an ambiguity in the conformal factor of the DeWitt metric, that is not fixed classically. These issues are subleading in the semiclassical regime that can be reliably discussed. A simple procedure is to quantise the constraint as written in e.g.~(\ref{eq:pipi}), with a flat DeWitt metric. One can then transform the quantum wave equation to hyperbolic coordinates. This is equivalent to writing down a covariant wave equation using the DeWitt metric. Following this procedure, the general solution is
\be\label{eq:psit}
\Psi(\tau,x_1,x_2,y) = \sum_k c_k \psi_k(x_1, x_2, y) \, e^{- \tau} e^{i \vep_k \tau} \,,
\ee
where the $c_k$ are coefficients and each mode obeys
\be\label{eq:modes}
- \nabla^2_{\H_3} \psi_k = - \Big[ y^2 \left( \partial_{x_1}^2 + \partial_{x_2}^2 + \partial_y^2 \right) - y \partial_y \Big] \psi_k = \left(1+ \vep_k^2\right) \psi_k \,.
\ee
See \cite{DeClerck:2023fax} for further discussion of (\ref{eq:psit}), including the exponentially decaying term and the choice of sign of the oscillating term. Our focus in the remainder is on the eigenmodes (\ref{eq:modes}).

The wave equation (\ref{eq:modes}) is the quantum analogue of the free motion Hamiltonian constraint (\ref{eq:Hamiltonian}). The analogue of the bounces is that the wavefunction must vanish at the walls,
\be\label{eq:bdy}
\psi_k \Big|_\text{walls} = 0 \,.
\ee
With these boundary conditions there will be a discrete spectrum of allowed $\vep_k$. We will see in \S\ref{sec:bianchi} that these are Maa{\ss} cusp forms. A potential continuous spectrum from non-holomorphic Eisenstein series is absent as these are not compatible with the boundary conditions.

We have seen in (\ref{eq:bdy:A2}), (\ref{eq:bdy:G2}) and (\ref{eq:bdy:C2}) that only one of the constraints involves the $y$ coordinate, and the remaining three define a triangle in the $(x_1,x_2)$ plane. This fact enables a further mode expansion, wherein each $\psi_k$ is decomposed in terms of modes that are supported in the allowed triangles, as we now explain. We will drop the $k$ label in the remainder, it is understood that we are considering a single `energy level' at a time. Then, let us write
\be\label{eq:Maass}
\psi(x_1,x_2,y) = \sum_{\lambda \geq 0} a_\lambda y K_{i\vep}(\l y) \, f_\l(x_1, x_2) \,,
\ee
where $K_{i\vep}(\l y)$ is a modified Bessel function that vanishes towards large $y$ and the modes obey
\begin{equation}\label{eq:feqn}
    - \left(\partial_{x_1}^2 + \partial_{x_2}^2 \right) f_\lambda(x_1,x_2) = \lambda^2 f_\lambda(x_1,x_2) \,.
\end{equation}
The allowed values of $\lambda$ will be determined below, by imposing that these modes vanish on the boundary of the appropriate $(x_1,x_2)$ plane triangle. Finally, the coefficients $a_\lambda$ are determined by imposing vanishing at the remaining wall,
\begin{equation}\label{eq:circle}
    \psi(x_1, x_2, y)=0, \qquad\text{where} \qquad x_1^2+x^2_2+y^2=1 \,.
\end{equation}
As is familiar from the conventional Schr\"odinger equation, the coefficients $c_k$ in the energy level expansion (\ref{eq:psit}) will be determined by initial conditions, while the `Fourier' coefficients $a_\lambda$ of the individual levels (\ref{eq:Maass}) are fixed by boundary conditions.

The three triangles of Fig.~\ref{fig:triangle} have the nontrivial property that the Dirichlet eigenmodes (\ref{eq:feqn}) are given by a finite superposition of trigonometric functions, see e.g.~\cite{brian}. This is ultimately due to their close relation to the three two dimensional algebras $C_2$, $G_2$ and $A_2$. We can consider the Gaussian and Eisenstein cases separately.

\subsection{The Gaussian domain}
\label{sec:gaussdom}

We will discuss the isosceles right-angled triangle in Fig.~\ref{fig:triangle} first, as this is the simplest case. The $(x_1,x_2)$ plane modes are clearly
\be\label{eq:fgau}
f_\lambda(z) = \sin(2\pi m x_1) \sin(2\pi n x_2) - \sin(2\pi m x_2) \sin(2\pi n x_1) \,,
\ee
with $\lambda^2 = (2 \pi)^2 (m^2 + n^2)$ for positive integers $m,n$ and $m \neq n$. We obtain (\ref{eq:fgau}) by solving the problem on the $[0,\half] \times [0,\half]$ square and then imposing vanishing along the diagonal $x_1 = x_2$.

The modes in (\ref{eq:fgau}) have the following symmetries
\be\label{eq:fsym}
f(z) = f(z+1) = f(z+i) = f(i z) = - f(\bar z) \,.
\ee
The first three of these transformations generate the symmetry group of the Gaussian integer lattice, given by translations and 90 degree rotations, as noted below (\ref{eq:gaussian}). These can be used to extend any function from the $[0,\half] \times [0,\half]$ square to the entire plane. The final transformation in (\ref{eq:fsym}) imposes vanishing along the lower boundary of this square. Combining this reflection with the generators of the symmetry group not only imposes vanishing on all boundaries of the square, but also along the diagonal $x_1 = x_2$. This last fact follows from $f(z) = - f(i \bar z)$. We therefore see that the right-angle triangle in Fig.~\ref{fig:triangle} is the natural domain for a Dirichlet problem associated to the Gaussian integers.

It will be convenient to write (\ref{eq:fgau}) in the form
\be\label{eq:fb}
f_\lambda(z) = \sum_{2 \pi |\mu| = \lambda} b_\mu e^{2 \pi i \langle \mu, i z \rangle} \,,
\ee
where $z = x_1 + i x_2$ (the function is not holomorphic in $z$) and the sum is over a finite set of complex numbers $\mu = \mu_1 + i \mu_2$ that have a fixed modulus. The bracket is defined as
\be\label{eq:brack}
\langle \mu, z \rangle \equiv \text{Re} \, \mu z = \mu_1 x_1 - \mu_2 x_2 \,. 
\ee
The minus sign in this bracket will be convenient for writing the Hecke operators later. The factor of $i$ in the bracket in (\ref{eq:fb}) will allow a unified discussion of the Gaussian and Eisenstein cases. As rotation by $i$ is a symmetry in (\ref{eq:fsym}), this factor is innocuous. We may now see how the symmetry conditions (\ref{eq:fsym}) fully determine the modes, up to normalisation. The translational symmetries require $\mu_1$ and $\mu_2$ to be integer. At a fixed $\lambda$ this allows for eight independent coefficients in (\ref{eq:fb}), which we can enumerate as follows. Letting $\mu_0 = m + i n$, the possible labels are $\mu = i^j \mu_0$ and $\mu = i^j \bar \mu_0$, with $j=0,1,2,3$. The other symmetries then imply that
\be\label{eq:bsym}
b_\mu = b_{i \mu} = b_{-\mu} = b_{-i \mu} = - b_{-\bar \mu} = - b_{-i \bar \mu} = - b_{\bar \mu} = - b_{i \bar \mu} \,.
\ee
These seven relations fix all eight coefficients in terms of, for example, $b_{\mu_0}$. The expression (\ref{eq:fb}) then reproduces (\ref{eq:fgau}), up to an overall normalisation.

Assembling the different modes into the full waveform we may therefore write
\be\label{eq:fullGauss}
\psi(z,y) = \sum_{\mu \in \Z[i]} b_\mu \, y \, K_{i\vep}(2 \pi |\mu|y) e^{2 \pi i \langle \mu, i z\rangle} \,.
\ee
Here $\psi(z,y)$ must obey (\ref{eq:fsym}). Equivalently, the $b_\mu$ coefficients must obey (\ref{eq:bsym}). The sum in (\ref{eq:fullGauss}) is over all Gaussian integers $m + i n$. Cases with $m = n$ or with one of $m$ or $n$ equal to zero vanish due to the symmetries. The unfixed $b_{m+in}$ coefficients must be determined by the remaining Dirichlet condition (\ref{eq:circle}). We will see shortly that this is equivalent to demanding modular invariance of $\psi(z,y)$ over $\Z[i]$.

\subsection{The Eisenstein domain}
\label{sec:eisdom}

\begin{figure}[h]
\centering
\includegraphics[width=0.4\textwidth]{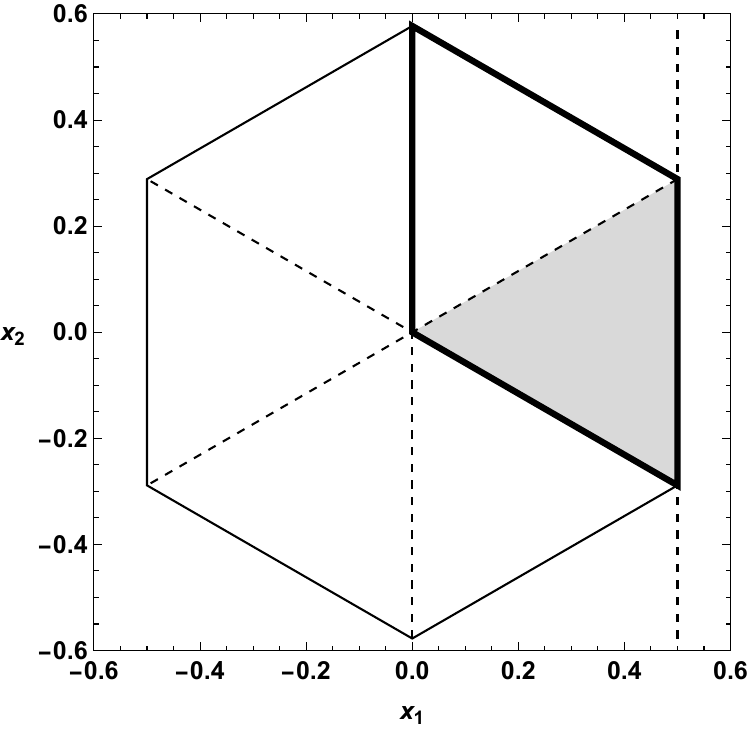}
\caption{A function defined in the shaded equilateral triangle and vanishing on the boundaries of the triangle can be extended to the entire plane using the transformations (\ref{eq:fsymE}). A reflection extends the function to the rhombus shown, Eisenstein rotations extend it to the hexagon shown, and Eisenstein  translations will then extend it to the entire plane. The function vanishes along the dashed lines shown.
\label{fig:hex}}
\end{figure}

We noted below (\ref{eq:eisenstein}) that the equilateral triangle in Fig.~\ref{fig:triangle} is closely related to the Eisenstein integers. For these integers, the analogous symmetries to (\ref{eq:fsym}) are
\be\label{eq:fsymE}
f(z) = f(z+1) = f(z+\omega) = f(\omega z) = - f(1 - \bar z) \,.
\ee
The first three of these transformations are symmetries of the Eisenstein lattice (these symmetries also include the transformation $z \mapsto -z$, discussed in \S\ref{sec:bianchi} below). The two translational symmetries can be used to map all points in the plane into the hexagon shown in Fig.~\ref{fig:hex}.
The rotational symmetry can then be used to further map the points into a rhombus, also shown 
in the figure. Conversely, the first three transformations can be used to extend any function from the rhombus to the entire plane. The final transformation in (\ref{eq:fsymE}) is a reflection that imposes vanishing of the function along the boundary of the rhombus at $x_1 = \frac{1}{2}$. Combining this reflection with a translation gives $f(z) = - f(-\bar z)$, and hence the function must also vanish along $x_1 = 0$. Rotations by $\omega$ therefore imply that the function furthermore vanishes along the lines $\text{arg}\, z = \pm \frac{\pi}{6}$. A function obeying (\ref{eq:fsymE}) therefore vanishes on the three boundaries of the equilateral triangle in Fig.~\ref{fig:triangle}, which is thus seen to be the natural domain for the Dirichlet problem associated to the Eisenstein integers.

As in the Gaussian case, we may look for a trigonometric solution to the constraints in (\ref{eq:fsymE}) in a form analogous to (\ref{eq:fb}). The translational invariance conditions in (\ref{eq:fsymE}) imply that the `momenta' must live in the dual lattice
\be\label{eq:ZZ}
\widehat{\Z[\omega]} = \frac{4 \pi i}{\sqrt{3}} \Z[\omega] \,.
\ee
Writing a general element of $Z[\omega]$ as $m + \omega n$ and then taking the norm of the the dual lattice elements in (\ref{eq:ZZ}), the eigenvalues in (\ref{eq:feqn}) are seen to be
\be
\lambda^2 = \frac{16 \pi^2}{3} \left(m^2 + n^2 - m n \right) \,.
\ee
Combining all of these modes together as in (\ref{eq:fullGauss}) we now obtain
\be\label{eq:fullEis}
\psi(z,y) = \sum_{\mu \in \Z[\omega]} b_\mu \, y \, K_{i\vep}({\textstyle \frac{4\pi}{\sqrt{3}}} |\mu|y) e^{\frac{4\pi}{\sqrt{3}} i \langle \mu, i z\rangle} \,.
\ee
Here the bracket is as defined previously in (\ref{eq:brack}) and the $i$ in the bracket is natural given (\ref{eq:ZZ}). At a fixed eigenvalue $\lambda^2$ there are now twelve independent coefficients. These are labeled by $\mu = (-\w)^j \mu_0$ and $\mu = (-\w)^j\bar{\mu}_0$, with $j={0,1,2,3,4,5}$ and where now $\mu_0=m+\omega n$. The remaining symmetries in (\ref{eq:fsymE}) imply that\footnote{It may be useful to recall at this point that the Eisenstein integers have six units $\{\pm 1, \pm \omega, \pm \omega^2\}$.}
\begin{align}\label{eq:bEis}
b_\mu  = b_{\omega \mu} = b_{\omega^2 \mu} = - b_{ \bar \mu} = - b_{ \omega \bar \mu} = - b_{ \omega^2 \bar \mu}   \,.
\end{align}
These five relations reduce the twelve independent coefficients down to two, for example $b_{\mu_0}$ and $b_{- \bar \mu_0}$. In contrast to the Gaussian case, then, the symmetries do not fully fix the mode.

The remaining modes can be classified by their sign under a further reflection
\be\label{eq:signs}
f(z) = \pm f(\bar z) \quad \Rightarrow \quad b_\mu = \pm b_{-\bar \mu} \,.
\ee
The choice of a definite sign here fully fixes the mode, up to normalisation. The explicit modes, written in terms of trigonometric functions, are given in \cite{brian}. For the 
hemi-equilateral triangle in Fig.~\ref{fig:triangle} we must choose the odd mode in order to impose Dirichlet boundary conditions along the real axis. For the equilaterial triangle in Fig.~\ref{fig:triangle}, however, all required Dirichlet boundary conditions are already in place from (\ref{eq:fsymE}). Therefore we may keep both modes in (\ref{eq:signs}) and the eigenvalue associated to a given $\mu_0$ is doubly degenerate. The symmetric wavefunction vanishes if and only if at least one of $m, n$ or $m-n$ is zero. The antisymmetric wavefunction vanishes if and only if at least one of $m,n$ or $m-n$ is zero, or if $m=2n, n=2m$ or $m+n=0$.

\subsection{Bianchi groups for the billiard domains}
\label{sec:bianchi}

To understand the symmetries in (\ref{eq:fsym}) and (\ref{eq:fsymE}) in terms of a modular group, we should describe the action of $PSL(2,\C)$ on the upper half space model of $\H_3$. It is convenient to package the $(z,y)$ coordinates into a hermitian matrix as
\be
M \equiv \frac{1}{y} \left(
\begin{array}{cc}
y^2 + |z|^2 & z \\
\bar z & 1
\end{array}
\right) \,.
\ee
The hyperbolic metric (\ref{eq:H32}) can then be written as
\be
ds^2 = \frac{1}{2} \Tr \left(M^{-1} dM M^{-1} dM \right) \,.
\ee
This expression is clearly invariant under
\be\label{eq:conj}
M \mapsto \gamma M \gamma^\dagger \,,
\ee
for any $\gamma \in GL(2,\C)$. This action preserves the hermiticity of $M$. However, in order to preserve the determinant of $M$ we require in addition that $\det \gamma = 1$, so that
\be
\gamma = 
\left(
\begin{array}{cc}
a & b \\
c & d
\end{array}
\right) \,, \qquad ad - bc = 1 \,.
\ee
Thus (\ref{eq:conj}) defines an action
of $PSL(2,\C)$ on $\H_3$. We may write this very explicitly as
\be\label{eq:expl}
(z,y) \mapsto \left(\frac{(az+b)(\bar c \bar z + \bar d) + a \bar c y^2}{|c z + d|^2 + |c|^2 y^2} , \frac{y}{|c z + d|^2 + |c|^2 y^2} \right) \,. 
\ee
At the conformal boundary $y=0$, the map (\ref{eq:expl}) becomes a M\"obius transformation
\be
z \mapsto \frac{a z + b}{c z + d} \,.
\ee

Using (\ref{eq:expl}), the translational symmetries in (\ref{eq:fsym}) and (\ref{eq:fsymE}) can be recognised as the $PSL(2,\C)$ transformations
\be
T_X \equiv \left(
\begin{array}{cc}
1 & X \\
0 & 1
\end{array}
\right) \,, \qquad (z,y) \mapsto (z + X,y) \,.
\ee
where $X=1,i$ in (\ref{eq:fsym}) and $X=1,\omega$ in (\ref{eq:fsymE}). The rotational symmetry in (\ref{eq:fsymE}) can be obtained as the $PSL(2,\C)$ transformation
\be
L_\omega \equiv \left(
\begin{array}{cc}
\omega^2 & 0 \\
0 & \omega
\end{array}
\right) \,, \qquad (z,y) \mapsto (\omega z,y) \,,
\ee
while {\it twice} the rotational symmetry in (\ref{eq:fsym}), i.e.~a rotation by 180 rather than 90 degrees, is obtained from the $PSL(2,\C)$ transformation
\be\label{eq:double}
L_{i} \equiv \left(
\begin{array}{cc}
-i & 0 \\
0 & i
\end{array}
\right) \,, \qquad (z,y) \mapsto (- z,y) \,.
\ee
We will return to this point shortly.

The most important step we will take in this section is to incorporate inversions, which we have not discussed yet. The inversion transformation $S \in PSL(2,\C)$ is, using (\ref{eq:expl}),
\be
S \equiv \left(
\begin{array}{cc}
0 & -1 \\
1 & 0
\end{array}
\right) \,, \qquad (z,y) \mapsto \left(\frac{- \bar z}{|z|^2 + y^2} , \frac{y}{|z|^2 + y^2} \right) \,. 
\ee
This transformation sends
\be\label{eq:inv}
|z|^2 + y^2 \mapsto \frac{1}{|z|^2 + y^2} \,,
\ee
and therefore leaves the locus $|z|^2 + y^2 = 1$ invariant. We recall from (\ref{eq:circle}) that the wavefunction must vanish on this locus. More immediately, however, we can use (\ref{eq:inv}) to map any point in $\H_3$ into a point with $|z|^2 + y^2 \geq 1$, which can then be inside the billiard domain.

The elements of $PSL(2,\C)$ just discussed generate the modular groups (see e.g.~\cite{fine, steil1996eigenvalues})
\be
PSL(2,\Z[i]) = \langle S, T_1, T_i, L_{i} \rangle \,, \qquad
PSL(2,\Z[\omega]) = \langle S, T_1, T_\omega, L_\omega \rangle \,.
\ee
The generators listed above obey relations following from their matrix definitions. These modular groups are both Euclidean Bianchi groups because $\Z[i]$ and $\Z[\omega]$ are the rings of integers in $\Q(\sqrt{-1})$ and $\Q(\sqrt{-3})$, respectively. These integers admit a Euclidean algorithm, a fact we will make use of in later sections. The group $PSL(2,\Z[i])$ is also called the Picard group.

The fundamental domains of the modular groups we have just described are plotted in e.g.~\cite{steil1996eigenvalues}. We can consider the two cases in turn. The fundamental domain of $PSL(2,\Z[\omega])$ is made up of all points in the rhombus in Fig.~\ref{fig:hex}, extended in the $y$ direction but with $y$ constrained to be above the `dome' $|z|^2 + y^2 = 1$. The billiard domains (\ref{eq:bdy:A2}) and (\ref{eq:bdy:G2}) are a little smaller than this fundamental domain. However, we have understood how this works out in the previous \S\ref{sec:eisdom}. To impose that the wavefunction vanish on all sides of the equilateral triangle we require oddness under the reflection $z \to - \bar z$. It follows that  wavefunctions in the equilateral Eisenstein billiard domain are equivalent to $\Z[\omega]$-automorphic functions on $\H_3$ that are odd under this reflection symmetry. That is, we have established that the WDW quantum billiard wavefunctions for the pure gravity theory 
in \S\ref{sec:E}  can be characterised as:
\be\label{eq:aa}
\text{5d gravity:} \quad \left\{
\begin{array}{cl}
\psi(z,y) & = \; \psi(\gamma(z,y)) \\
\psi(z,y) & = \; - \psi(-\bar z,y) 
\end{array} \right.\,, \quad \gamma \in PSL(2,\Z[\omega]) \,.
\ee
Furthermore, the wavefunctions (\ref{eq:fullEis}) decay towards the cusp at $y \to \infty$, and therefore define odd Maa{\ss} cusp forms.

We also saw in \S\ref{sec:eisdom} that to impose vanishing on all sides of the hemi-equilateral triangle we furthermore require oddness under $z \to \bar z$. We could directly impose a further reflection symmetry, but a nicer way to reduce the domain was given in \cite{Feingold:2008ih}. The modular group $PSL(2,\Z[\omega])$ admits an index two extension by the $PSL(2,\C)$ transformation 
\be\label{eq:ayi}
A \equiv \left(
\begin{array}{cc}
- i \omega & 0 \\
0 & i  \omega^2
\end{array}
\right) \,, \qquad (z, y) \mapsto (- \omega^2 z, y) \,,
\ee
which is not in $PSL(2,\Z[\omega])$. It is shown in \cite{Feingold:2008ih} that the extended group is the semi-direct product $PSL(2,\Z[\omega]) \rtimes 2$ (our $A = S S_2$, in the notation of \cite{Feingold:2008ih}). Combining (\ref{eq:ayi}) with a rotation by $\omega$ gives the symmetry transformation $z \to - z$. Oddness under the reflection $z \to -\bar z$ then implies oddness under $z \to \bar z$, as desired. It follows that the WDW quantum billiard wavefunctions for the Einstein-Maxwell theory in \S\ref{sec:EM} are given by the odd Maa{\ss} forms:
\be\label{eq:bb}
\text{5d Einstein-Maxwell:} \quad \left\{
\begin{array}{cl}
\psi(z,y) & = \; \psi(\gamma(z,y)) \\
\psi(z,y) & = \; - \psi(-\bar z,y) 
\end{array} \right. \,, \quad \gamma \in PSL(2,\Z[\omega]) \rtimes 2 \,.
\ee

The fundamental domain of $PSL(2,\Z[i])$ is made up of the rectangle $[0,\frac{1}{2}]\times[-\frac{1}{2},\frac{1}{2}]$ in the $z$ plane, extended in the $y$ direction above the `dome' $|z|^2 + y^2 = 1$. This is, again, larger than the Gaussian billiard domain (\ref{eq:bdy:G2}).
Imposing oddness under the reflection $z \to \bar z$, as in (\ref{eq:fsym}), will reduce the domain down to $[0,\frac{1}{2}]\times[0,\frac{1}{2}]$. However, a further reduction is needed
in order to obtain the isosceles right-angled triangle in Fig.~\ref{fig:triangle}. The issue is that in (\ref{eq:fsym}) we furthermore imposed invariance under rotations by $i$, but (\ref{eq:double}) shows that only rotations by twice that amount are in $PSL(2,\Z[i])$. The rotation that we need is instead the $PSL(2,\C)$ transformation
\be\label{eq:byi}
B \equiv \left(
\begin{array}{cc}
e^{i \pi/4} & 0 \\
0 & e^{-i \pi/4}
\end{array}
\right) \,, \qquad (z, y) \mapsto (i z, y) \,,
\ee
which is not in $PSL(2,\Z[i])$. However, it is shown in \cite{Feingold:2008ih} that $B$ can be added to the modular group, giving an index two extension of the group to $PSL(2,\Z[i]) \rtimes 2$. Therefore, the WDW quantum billiard wavefunctions for the EMAD theory in \S\ref{sec:EMAD} are odd Maa{\ss} forms obeying
\be\label{eq:4dfin}
\text{4d EMAD:} \quad \left\{
\begin{array}{cl}
\psi(z,y) & = \; \psi(\gamma(z,y)) \\
\psi(z,y) & = \; - \psi(\bar z,y) 
\end{array} \right.\,, \quad \gamma \in PSL(2,\Z[i]) \rtimes 2 \,.
\ee
Because $\psi(z,y) = \psi(-z,y)$ from the first line of (\ref{eq:4dfin}), we could have  equivalently imposed $\psi(z,y) = - \psi(-\bar z,y)$ in the second line of (\ref{eq:4dfin}). That choice of reflection would emphasise the similarity with the previous cases (\ref{eq:aa}) and (\ref{eq:bb}).

We end this section with two brief comments. Firstly, as we mentioned already, the extensions of the modular groups discussed above can equivalently be thought of as an additional choice of even- or odd-ness within the usual modular group. This is the perspective taken in \cite{steil1996eigenvalues}. Secondly, in addition to the Maa{\ss} forms there is the non-holomorphic Eisenstein series which has a continuous spectrum. However, the Eisenstein series is even under $z \mapsto - \bar z$ and therefore does not satisfy our Dirichlet boundary conditions.

\section{Dilatations, Rotations and $L$-functions}
\label{sec:Lfun}

In this section we will Mellin transform
the automorphic WDW wavefunctions of \S\ref{sec:auto} into a sequence of $L$-functions. This is the result advertised in (\ref{eq:map3}) above. There is an extra step compared to the case of four-dimensional gravity \cite{Hartnoll:2025hly} because of the larger symmetry group of $\H_3$ compared to $\H_2$: In addition to a Mellin transform in the radial direction, we must Fourier decompose in an angular direction. This produces an $L$-function for each angular momentum mode. In \S\ref{sec:primon} we will construct the primon gas partition functions for these $L$-functions.

\subsection{Symmetry algebra}\label{sec:UHS}

The Lie algebra of infinitesimal isometries of $\H_3$ is
$\mathfrak{so}(3,1) \cong \mathfrak{sl}(2,\mathbb{C})$. This is a six dimensional algebra generated by two translations
\begin{equation}
        T_1 = \partial_{x_1} \,, \qquad T_2 = \partial_{x_2} \,,
\end{equation}
a rotation and a dilatation
\begin{equation}
        R = x_1 \partial_{x_2} - x_2 \partial_{x_1} \,, \qquad D = x_1 \partial_{x_1} + x_2 \partial_{x_2} + y \partial_y \,,
\end{equation}
and two generators of special conformal transformations
\begin{align}
        K_1 &= (x_2^2 - x_1^2 + y^2) \partial_{x_1} - 2x_1 x_2 \partial_{x_2} - 2x_1 y \partial_y, \\
        K_2 &= (x_1^2 - x_2^2 + y^2) \partial_{x_2} - 2x_2 x_1 \partial_{x_1} - 2x_2 y \partial_y \,.
\end{align}
It is immediately noted that
\be
[R,D] = 0 \,,
\ee
so that states can have simultaneous angular momentum and dilatation eigenvalues.

The Casimir of this algebra is precisely the Laplacian (\ref{eq:modes}) on $\H_3$:
\be
C_2 \equiv K_1 T_1 + K_2 T_2 + D(D-2) - R^2 = y^2 \left(\pa_{x_1}^2 + \pa_{x_2}^2 + \pa_y^2 \right) - y \pa_y \,.
\ee
It follows that the automorphic state $\psi$ is in a representation of $\mathfrak{sl}(2,\mathbb{C})$ with $C_2 = - (1 + \vep^2)$. The representation can be identified precisely by constructing the two commuting copies of $\mathfrak{sl}(2,\R)$ inside $\mathfrak{sl}(2,\mathbb{C})$: $K_\pm = K_1 \pm i K_2$, $T_\pm = T_1 \mp i T_2$, $D_\pm = D \mp i R$. The $\mathfrak{sl}(2,\R)$ generators define Casimirs $C_{2\pm}$. One finds that $C_2 = C_{2+} = C_{2-}$. The equality of the $\mathfrak{sl}(2,\R)$ Casimirs combined with $C_2 = - (1 + \vep^2)$ corresponds to a particular principal series representation of $\mathfrak{sl}(2,\mathbb{C})$ where a certain discrete label vanishes and a certain continuous label $s= \vep$, see e.g.~\cite{taylor1986noncommutative}. As in \cite{Hartnoll:2025hly}, the boundary scaling dimension associated to this (unitary) representation will be complex.
That is, as $y \to 0$ solutions to the Laplace equation behave as $y^{1 \pm i \vep}$.

It will be helpful to introduce spherical coordinates $(\rho, \vartheta, \varphi)$ on $\H_3$, defined as
\begin{equation}\label{eq:polar}
     (x_1, x_2, y) \equiv \rho(\sin\vartheta\cos\varphi, \sin\vartheta\sin\varphi, \cos\vartheta) \,.
\end{equation}
In these coordinates the rotation and dilatation operators simplify to
\be\label{eq:RD}
R = \pa_{\varphi} \,, \qquad D = \rho \, \pa_\rho \,.
\ee

\subsection{Fourier and Mellin decompositions}
\label{sec:FM}

In this subsection we are going to express the Maa{\ss} waveforms
\be\label{eq:maass}
\psi(z,y) = \sum_{\mu \in \ocal} b_\mu \, y \, K_{i\vep}(\alpha |\mu|y) e^{i \alpha \langle \mu,  i z\rangle} \,,
\ee
obtained in (\ref{eq:fullGauss}) and (\ref{eq:fullEis}), as a sum over rotation and dilatation eigenstates. The value of $\alpha$ in (\ref{eq:maass}) depends on the integer lattice $\ocal$: For Gaussian integers, $ \alpha = 2\pi $, while for Eisenstein integers, $ \alpha = \frac{4\pi}{\sqrt{3}} $. Finally, recall that $z = x_1 + i x_2$.

The Jacobi–Anger expansion expresses the phase in (\ref{eq:maass}) in terms of Bessel functions,
\begin{equation}\label{eq:JA}
    e^{i \a \langle \mu, i z \rangle} = \sum_{m=-\infty}^{\infty} (-1)^m J_m(\a \,  |\mu| \rho \sin \vartheta) \, e^{i m (\varphi + \arg(\mu))} \,.
\end{equation}
Here we used the polar coordinates (\ref{eq:polar}) and defined the argument by
\be
\mu_1 + i \mu_2 \equiv|\mu| e^{i \arg(\mu)} \,.
\ee
From (\ref{eq:JA}) we obtain the Fourier modes in $\varphi$ of the Maa{\ss} waveform (\ref{eq:maass}) as
\begin{equation}
\begin{aligned}
    \psi_n(\rho, \vartheta) 
    & \equiv \int_0^{2\pi} \psi(z,y) e^{i n \varphi} \, d\varphi \\
    & = 2 \pi (-1)^n \rho \cos\vartheta \,\sum_{\mu \in \ocal} b_\mu \,  K_{i\vep}(\alpha |\mu| \rho \cos\vartheta)
    J_n(\a |\mu| \rho \sin \vartheta) \,  e^{- i n \arg(\mu)} \,. \label{eq:psin}
\end{aligned}
\end{equation}
From (\ref{eq:RD}), this is equivalent to expressing the waveform in a basis of rotation eigenstates. The periodicity of the Maa{\ss} form depends on the symmetry of the underlying triangle. Accounting for the reflection symmetries across the edges of the triangle, waveforms corresponding to the three triangles in Fig.~\ref{fig:triangle} have angular periodicity $\frac{2\pi}{3}, \frac{\pi}{3}$ and $\frac{\pi}{2}$. The corresponding $\psi_n$s are therefore only nonzero for $n$ divisible by $3, 6$ and $4$, respectively.

We may now furthermore perform a Mellin transform on the Fourier modes (\ref{eq:psin}), thereby expressing the wavefunction in a simultaneous basis of rotation and dilatation eigenstates. From (\ref{eq:RD}), the delta function normalisable dilatation eigenstates are $\rho^{-i t}$. Therefore we define
\begin{equation}
\begin{aligned}
    M_n(t, \vartheta) 
    & \equiv \int_0^\infty \frac{d\rho}{\rho} \, \rho^{i t} \, \psi_n(\rho, \vartheta) \\
    &= L^-_n(1 + i t)\, F_n(t,\vartheta) \,, \label{eq:LL}
\end{aligned}
\end{equation}
where we define the $L$-function (the label $-$ will be useful later) 
\begin{equation}\label{eq:Lminus}
    L^-_n(s) \equiv \sum_{\mu \in \mathcal{O}} \frac{b_\mu \, e^{-in\arg(\mu)}}{|\mu|^s} \,,
\end{equation}
and the angular part is given in terms of a hypergeometric function, note that $F_n=F_{-n}$,
\begin{equation}
\begin{aligned}
    F_n(t, \vartheta) \equiv & \frac{2 \pi (-1)^n}{\a^{1 + i t} (\cos\vartheta)^{i t}} \int_0^\infty d\bar \rho \, \bar \rho^{it} \,  K_{i\vep}(\bar \rho)
    J_n(\bar \rho \tan \vartheta) 
    \\
    = & \frac{\pi}{|n|! \a} \left({\textstyle \frac{2}{\a}} \sec\vartheta\right)^{i t} \tan^{|n|} (-\vartheta) \, \Gamma\left(\half \left[i( t - \vep) + |n| + 1\right] \right) \Gamma\left( \half \left[i(t + \vep) + |n| + 1\right] \right) \\
    &\quad \times {}_2F_1\left( \half \left[i( t - \vep) + |n| + 1\right], \half \left[i(t + \vep) + |n| + 1\right], |n| + 1, -\tan^2 \vartheta \right).
\end{aligned}
\end{equation}
Here $\bar \rho$ is a rescaling of $\rho$ such that in the second line of (\ref{eq:LL}) we have been able to separate out the sum over $\mu$ from the integral.

Under the inversion map (\ref{eq:inv}) followed by the reflection $z \to - \bar z$, that imposes oddness of the Maa{\ss} forms, we have
\begin{equation}\label{eq:ref:psi}
    \psi(\rho, \vartheta, \varphi) = -\psi\left( \frac{1}{\rho}, \vartheta, \varphi \right) \,.
\end{equation}
If we use this transformation in (\ref{eq:LL}), we deduce that $M_n(t,\vartheta) = - M_n(-t,\vartheta)$.\footnote{The transformation (\ref{eq:ref:psi}) also implies that $M_n$ is an entire function of $t$, by re-writing the
integral in \eqref{eq:LL} as
\begin{equation}
    \int_1^\infty \frac{d\rho}{\rho} (\rho^{it}-\rho^{-it})\psi_n(\rho, \vartheta) \,.
\end{equation}
Given that $\psi_n(\rho,\vartheta)\sim e^{-\alpha|\mu|\cos\vartheta\rho}$ at large $\rho$, this integral is absolutely convergent for all complex $t$.}
This fact then gives
a reflection formula for the $L$-function, which is most elegantly stated as
\be\label{eq:refl}
\xi^-_n(s) = - \xi^-_n(2-s) \,,
\ee
where the xi function
\be\label{eq:xi}
\x^-_n(s) \equiv \left(\textstyle \frac{2}{\a}\right)^{s} \Gamma\left(\half \left[|n| + s - i \vep\right] \right) \Gamma\left( \half \left[|n| + s + i \vep\right] \right) L^-_n(s) \,.
\ee
In Appendix \ref{app:adelic} we explain how the xi function (\ref{eq:xi}) and its reflection formula (\ref{eq:refl}) are naturally interpreted in an adelic framework. Such reflection formulae are a defining feature of $L$-functions.

The result (\ref{eq:LL}) is the relation advertised in (\ref{eq:map3}). We have shown that the WDW wavefunctions in a basis of dilatation and rotation eigenstates are proportional to certain $L$-functions (\ref{eq:Lminus}) along the fixed line $\text{Re}(s) = 1$ of the reflection (\ref{eq:refl}). This line is where the nontrivial zeros of the function are expected to lie, according to the generalised Riemann hypothesis. This structure is very similar to the case of $SL(2,\Z)$ Maa{\ss} forms \cite{Hartnoll:2025hly}. The differences are firstly that the Dirichlet coefficients in the $L$-function (\ref{eq:Lminus}) are labeled by nontrivial integer rings $\ocal$ and secondly that an $n$-dependent phase multiplies the Dirichlet coefficients in (\ref{eq:Lminus}).

\section{Complex primon gases}
\label{sec:primon}

The logic so far has been as follows. In \S\ref{sec:five} we reviewed the billiard domains arising from the near-singularity dynamics of certain gravity theories. In
\S\ref{sec:auto} we reviewed how WDW eigenfunctions in these domains are given by automorphic Maa{\ss} forms on $\H_3$. We have just seen in \S\ref{sec:Lfun} that when expressed in a dilatation and angular momentum basis the eigenfunctions are $L$-functions evaluated along their critical axes.

In this section we will relate the $L$-function (\ref{eq:Lminus}) to a trace over an auxiliary Hilbert space of oscillators. This is the dual primon gas. The key step here is an Euler product formula. To be able to write this formula we will need to review prime factorisation in our rings of integers $\ocal$, and also the Hecke relations that allow all Dirichlet coefficients to be expressed in terms of prime coefficients, i.e.~those with $\mu \in {\mathcal P}_\ocal$. Our discussion here will closely follow \cite{steil1996eigenvalues}.

\subsection{Hecke relations}
\label{sec:hecrelate}

In this subsection we will explain how the $b_\mu$ coefficients in (\ref{eq:maass}) can be expressed in terms of the prime coefficients. There will be a complication with imposing oddness under the reflection $z \mapsto - \bar z$.
Therefore, to start with, we consider 
\be\label{eq:cc}
f(z,y) = \sum_{\mu \in \ocal} c_\mu \, y \, K_{i\vep}(\alpha |\mu|y) e^{i \alpha \langle \mu,  i z\rangle} \,,
\ee
which are invariant under $PSL(2,\ocal)$ only, without any parity requirement under $z \mapsto - \bar z$.\footnote{We will see below that the index two extensions of the modular groups, discussed in \S\ref{sec:bianchi}, are enforced by the Hecke relations together with choices for the unit coefficients.} To highlight this weaker invariance, we have denoted the coefficients by $c_\mu$ rather than $b_\mu$.

In Appendix \ref{app:heck} we recall the existence of Hecke operators. These operators are self-adjoint on $PSL(2,\mathcal{O})$-invariant functions, commuting amongst themselves and also with the Laplacian on $\mathbb{H}_3$. Therefore, there exists a complete set of joint eigenfunctions for the Laplacian and Hecke operators. Taking (\ref{eq:cc}) to be an eigenfunction of the Hecke operators one finds that the $c_\mu$ coefficients are the Hecke eigenvalues and must therefore be real. Furthermore, for any $\mu, \nu \in \mathcal{O}$, and in a convention where $c_1 = 1$, one obtains the Hecke relations:
\begin{equation}
    c_\mu c_\nu = \sideset{}{^\times}\sum_{d|(\mu,\nu)} c_{\frac{\mu\nu}{d^2}} \,.
    \label{eq: hecke relations}
\end{equation}
The sum here is over all common divisors of $\mu$ and $\nu$, modulo units (this is what the $\times$ indicates).

The relations (\ref{eq: hecke relations}) can be used to construct all coefficients out of the prime and unit coefficients. This is possible because both of our rings $\ocal$ are unique factorisation domains: non-zero and non-unit elements of these rings can be written uniquely as a product of primes in ${\mathcal P}_\ocal$, up to ordering and multiplication by units. That is, for all $\mu \in \ocal$ one can write
\be\label{eq:factor}
\mu = u \prod_i \cp_i^{\alpha_i} \,, \qquad \text{where} \qquad -\frac{\pi}{n_u} < \arg(\cp_i) \leq \frac{\pi}{n_u} \,.
\ee
where $u$ is a unit, the $\cp_i$ are complex primes and there are $n_u = |\ocal^\times|$ units in total. Here $\times$ denotes the units in a ring. The units allow the argument of the primes to be constrained to the range shown in (\ref{eq:factor}). Recall that the units for the Gaussian integers are $\Z[i]^\times =\{\pm 1, \pm i\}$ and units for the Eisenstein integers are $\Z[\omega]^\times = \{\pm 1, \pm \omega, \pm \omega^2\}$. Thus $n_u = 4$ and 6, respectively. We will characterise the primes for both rings shortly. Thus from (\ref{eq: hecke relations}) we have
\begin{align}
c_{\mu\nu} & = c_\mu c_\nu && \text{for} \quad (\mu,\nu) \in \ocal^\times \,,
\label{eq:h1} \\
c_{\cp^{n+1}} & = c_{\cp^n} c_\cp - c_{\cp^{n-1}} && \text{for} \quad \cp\in {\mathcal P}_\ocal \,. \label{eq:h2}
\end{align}
Here $(\mu,\nu)$ is the greatest common divisor.
The first of these relations allows a coefficient to be broken up, multiplicatively, into prime power and unit factors. The second relation allows the prime power coefficients to be expressed, recursively, in terms of prime coefficients. To give explicit examples we must describe the Gausssian and Eisenstein primes.

\begin{figure}[h]
\centering
\includegraphics[width=0.4\textwidth]{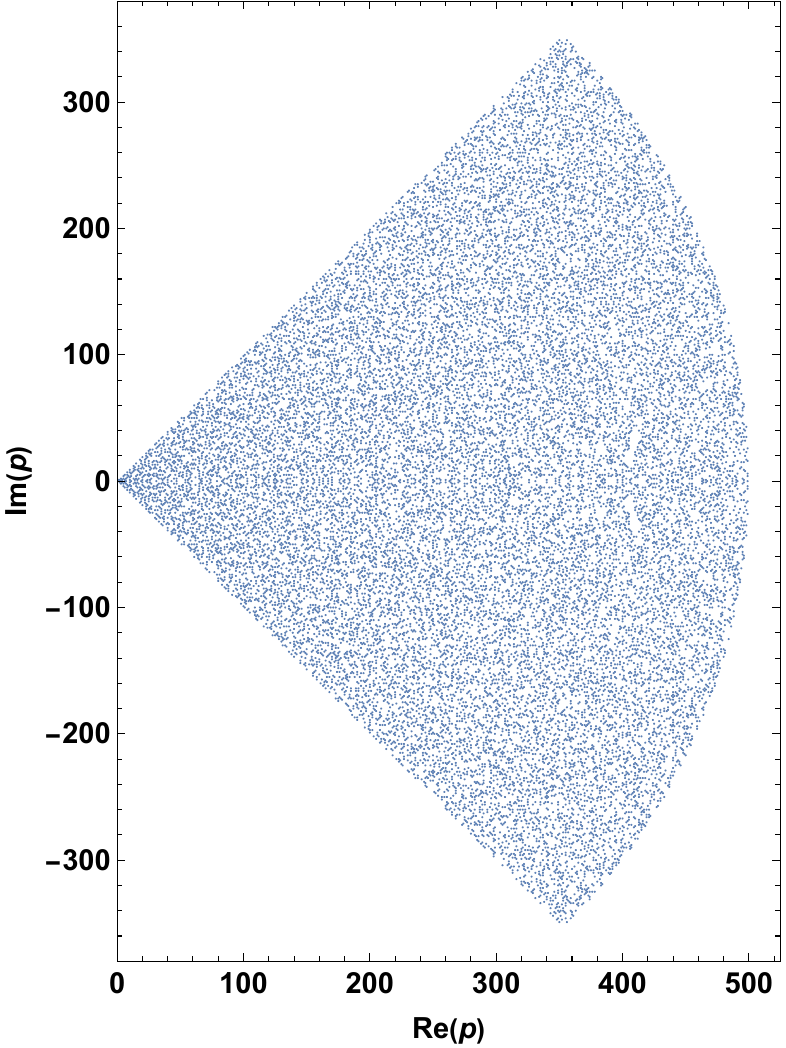}\hspace{1cm}
\includegraphics[width=0.4\textwidth]{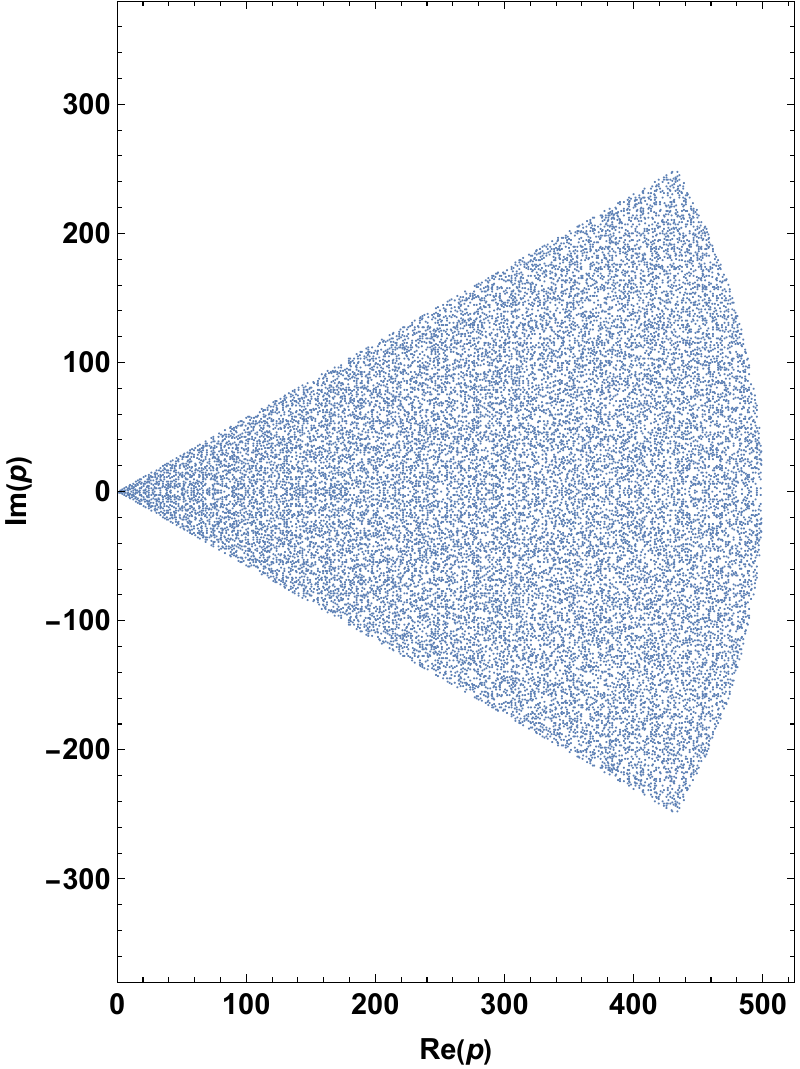}
\caption{Gaussian (left) and Eisenstein (right) primes with $|\cp| < 500$.}
\label{fig:gaussp}
\end{figure}

The prime Eisenstein and Gaussian integers can both be understood from the fate of the ordinary (`rational') primes in $\Z$ when embedded into these rings. The ordinary primes are either inert, ramify or split. That is, they either remain prime, can be written as the square of a prime or can be written as a product of two primes. This leads to different classes of primes in $\ocal$, as we now briefly review.
We can restrict to primes in the domain (\ref{eq:factor}).
Gaussian primes are of the form either {\it (a)} $m + i n$ with $m,n \neq 0$ and $m^2 + n^2$ a prime, which will be in $4 \N + 1$ or equal to 2, or {\it (b)} a prime $p \in 4 \N +3$. Case {\it (b)} comes from inert rational primes while case {\it (a)} comes from rational primes that split (those in $4 \N + 1$) or ramify (the case of $2 = - i (1+i)^2$). Thus, for example, in the Gaussian integer case we have
\be
c_{2} = c_{-i} \left(c_{1+i}^2 - 1\right) \,, \qquad c_{9 + 12 i} = c_{3} \left(c_{2+i}^2 - 1 \right) \,, \qquad c_{7 i} = c_i c_7 \,. \label{eq:exg}
\ee
All Gaussian primes $\cp$ with $|\cp| < 500$ are shown in Fig.~\ref{fig:gaussp}.
On the other hand, the Eisenstein primes are elements of the form either {\it (a)} $m + n \omega$ where $m^2 + n^2 - m n$ is a prime in $3 \N + 1$ or {\it (b)} a prime $p$ in $3 \N + 2$ or {\it (c)} $1-\omega$. These cases respectively correspond to rational primes that split, are inert or ramify (because $3 = -\omega^2(1-\omega)^2$). Thus, for example, in the Eisenstein integer case we have
\be
c_3 = c_{-\omega^2} (c^2_{1-\omega} - 1) \,, \qquad  c_{16 + 10\omega} = c_2\left(c_{3 + \omega}^2 - 1 \right) \,, \qquad c_{5 \omega} = c_\omega c_5 \,. \label{eq:exe}
\ee
All Eisenstein primes $\cp$ with $|\cp| < 500$ are shown in Fig.~\ref{fig:gaussp}.

The first Hecke relation (\ref{eq:h1}) applies, in particular, when one of $\mu$ or $\nu$ is a unit, as we saw in (\ref{eq:exg}) and (\ref{eq:exe}). There is some freedom in choosing the Hecke coefficients for the units. For the Gaussian case, invariance under rotation by $i$ forces us to take $c_\mu = 1$ for all units.
For the Eisenstein case the choice $c_\mu = 1$ for all units is again consistent with the symmetries and corresponds to the smaller triangle 
in Fig.~\ref{fig:triangle}. The associated Hecke eigenfunctions are then invariant under multiplication by any unit. The other choice allowed by the symmetries in (\ref{eq:bEis}) is $1 = c_{1} = c_{\omega} = c_{\omega^2} = - c_{-1} = - c_{-\omega} = - c_{-\omega^2}$. This choice with minus signs corresponds to the even sector of the equilateral triangle. The choice with all plus signs is the odd sector of the equilateral triangle in Fig.~\ref{fig:triangle}. The need to fix the unit Hecke coefficients gives a further perspective on the extension of the modular groups by (\ref{eq:ayi}) and (\ref{eq:byi}). The smaller triangle has a bigger symmetry group. We see, again, that the extension of the groups is equivalent to choosing a certain evenness property of the automorphic forms. See Fig.~3 in \cite{steil1996eigenvalues}.

The reflection $z \mapsto -\bar z$ is more problematic than the rotations we have just discussed. Unlike the case of $PSL(2,\mathbb{Z})$ Maa{\ss} forms \cite{Hartnoll:2025hly}, this reflection does not commute with the Hecke operators \cite{steil1996eigenvalues}.
In particular, odd Maa{\ss} forms cannot be eigenfunctions of the Hecke operators. This can be seen from the fact that either (\ref{eq:bsym}) or (\ref{eq:bEis}), for the Gaussian and Eisenstein cases respectively, imply that $b_1 = - b_1 = 0$. In contrast, the first Hecke eigenvalue $c_1 \neq 0$ (see Appendix \ref{app:heck}). However, we can construct odd and even Maa{\ss} forms by taking a linear superposition of the Hecke eigenforms (\ref{eq:cc})
\be\label{eq:pff}
\psi_\pm(z,y) = \half \left[f(z,y) \pm f(-\bar z,y) \right]\,.
\ee
The two Maa{\ss} forms constructed in this way have the same 
$\vep$ eigenvalue, although in cases where the Hecke eigenform is even then there is no corresponding odd Maa{\ss} form \cite{Aurich:2004ik}.
In terms of the coefficients, (\ref{eq:pff}) is equivalent to taking
\begin{equation}
    b^\pm_\mu = \frac{c_\mu \pm c_{\bar{\mu}}}{2} \,.
    \label{eq: relation an cn}
\end{equation}
Recall that we have set $c_1 = 1$. Our interest is in odd waveforms defined by the $b^-_\mu$. In Appendix \ref{app: hejhal} we see that (\ref{eq: relation an cn}) is invertible, and hence all odd waveforms come from antisymmetrising a Hecke eigenform. It will be the Hecke eigenforms that connect most directly to a primon gas.

Given a Hecke eigenform (\ref{eq:cc}) we can construct the $L$-functions
\begin{equation}\label{eq:Ltot}
    L_n(s) \equiv \sum_{\mu \in \mathcal{O}} \frac{c_\mu \, e^{-in\arg(\mu)}}{|\mu|^s} \,.
\end{equation}
The phase term here is an instance of a Hecke character \cite{neukirch1999} for the Gaussian and Eisenstein integers. From (\ref{eq: relation an cn}), this $L$-function is related to $L$-functions built from the $b_\mu^\pm$ by
\be\label{eq:LLL}
L_n^\pm(s) = \half \left[ L_{n}(s) \pm L_{-n}(s) \right]\,.
\ee
We encountered $L_n^-(s)$ previously in (\ref{eq:Lminus}). 
Recalling that the $c_\mu$ are real, for real $s$ we have
$L_n^-(s) = \text{Im} \, L_{n}(s)$. As we saw in \S\ref{sec:FM}, however, we are also interested in complex $s$. Our comment about invertibility, below (\ref{eq: relation an cn}), implies that every $L_n^{-}(s)$ comes from a pair $L_{\pm n}(s)$ according to (\ref{eq:LLL}). In the following \S\ref{sec:primon2} it will be the $L_{n}(s)$ that have a direct interpretation as a primon gas partition function. To write down the reflection law for the
$L_{n}(s)$ we can define a xi function $\xi_n(s)$ exactly as in (\ref{eq:xi}), just with $L^-_n(s) \to L_{n}(s)$. Then we have that
\be\label{eq:xipm}
\xi_{\pm n}(s) = \xi_{\mp n}(2-s) \,.
\ee
Note that $\xi_{\pm n}$ are exchanged under this reflection.

The $L$-functions (\ref{eq:Ltot}) are not independent, in the sense that all of the $c_\mu$ can be extracted from any one of them. Due to the Hecke relations, it is sufficient to extract the prime coefficients. A Mellin transform (called a Perron transform in this context) can be used to extract all of the terms in (\ref{eq:Ltot}) with a given magnitude $|\mu|$, i.e.~to obtain $\sum_{|\nu| = |\mu|} c_\nu e^{-in\arg(\nu)}$. When $\mu = \cp$ is a prime, the only possible $\nu$ with the same magnitude are $\cp$ or $\overline{\cp}$ multiplied by units. The unit factors can be turned into an overall constant using Hecke relations (cf.~(\ref{eq:AA}) below). This leaves at most a sum of two terms: $c_\cp \, e^{-in\arg(\cp)} + c_{\overline{\cp}} \, e^{i n\arg(\cp)}$. Because the Hecke eigenvalues are real, we can then determine $c_\cp$ and $c_{\overline{\cp}}$ uniquely from the value of this sum.

\subsection{Euler product and automorphic primon gas partition function}
\label{sec:primon2}

The $L_{n}(s)$ functions (\ref{eq:Ltot}) associated with Hecke eigenforms admit an Euler product representation
\begin{align}
L_n(s) & = A_n \prod_{\cp \in {\mathcal P}_\ocal} \sum_{m=0}^\infty \frac{c_{\cp^m}e^{- i n m \arg(\cp)}}{|\cp|^{sm}} \\
& = A_n \prod_{\cp \in {\mathcal P}_\ocal}
\frac{1}{1 - c_\cp\, e^{-in\arg(\cp)} |\cp|^{-s} + e^{-2in\arg(\cp)} |\cp|^{-2s}} \,.\label{eq:euler}
\end{align}
The first and second steps here follow from the two Hecke relations (\ref{eq:h1}) and (\ref{eq:h2}), respectively.

The prefactor due to the units is
\be\label{eq:AA}
A_n \equiv \sum_{u \in \ocal^\times} c_u \, e^{- i n \arg(u)} = \sum_{j=1}^{n_u} c_{u_0}^j \, e^{- i n j \arg(u_0)}  \,.
\ee
Here $u_0$ is a generator for the units, e.g. $u_0 = i$ for the Gaussians and $u_0 = - \omega$ for Eisenstein integers. We have used $c_{u_0^j} = c_{u_0}^j$, from (\ref{eq:h1}). Recall that the number of units $n_u$ is 4 for Gaussian and 6 for Eisenstein. We can evaluate the sum in (\ref{eq:AA}) in the three cases of interest. As discussed above, for the Gaussian case all $c_u = 1$ while for the Eisenstein case either again all $c_u = 1$ or $c_u = -c_{-u} = 1$, now with $u=1,\omega,\omega^2$. Thus we obtain that the only nonzero coeffients are, respectively,
\be\label{eq:Avals}
A^{\Z[i]}_{n \in 4 \Z}  = 4 \,, \qquad
A^{\Z[\omega],+}_{n \in 6 \Z} = 6 \,, \qquad
A^{\Z[\omega],-}_{n \in 3+6 \Z} =  6 \,.
\ee
The nonzero values of $n$ are the same as noted previously for the Maa{\ss} forms below (\ref{eq:psin}).

The Ramanujan conjecture that $|c_\cp| \leq 2$ is corroborated numerically \cite{steil1996eigenvalues, Aurich:2004ik}. With this assumption it is convenient to introduce angles $\theta_p$ such that
\be
c_\cp \equiv 2 \cos \theta_\cp \,.
\ee
In terms of these angles we can write (\ref{eq:euler}) as
\begin{align}\label{eq:euler2}
L_n(s) = A_n \prod_{\cp \in {\mathcal P}_\ocal}
\frac{1}{1 -  e^{-i(n\arg(\cp)-\theta_\cp)}|\cp|^{-s} }\frac{1}{1-e^{-i(n\arg(\cp)+\theta_\cp)} |\cp|^{-s}} \,.
\end{align}
As in \cite{Hartnoll:2025hly}, we can recognise the factors in (\ref{eq:euler2}) as partition functions of charged bosonic oscillators. This fact allows (\ref{eq:euler2}) to be interpreted as a trace over the Fock space of these oscillators, as we now explain.

For each prime $\cp \in {\mathcal P}_\ocal$, consider a pair of oscillators with creation operators $b_\cp^\dagger$ and $c_\cp^\dagger$. Let the Hamiltonian, charge and `angular momentum' of these oscillators be
\begin{align}
H_\cp & \equiv \log|\cp|
\left(b^\dagger_\cp b_\cp + c^\dagger_\cp c_\cp\right) \,, \\
Q_\cp & \equiv b^\dagger_\cp b_\cp-c^\dagger_\cp c_\cp \,, \\
L_\cp & \equiv \arg(\cp) \left(b^\dagger_\cp b_\cp+c^\dagger_\cp c_\cp\right) \,.
\end{align}
Then we have the complex primon gas partition function:
\be\label{eq:final}
L_n(s) = A_n \, \text{tr} \exp{\sum_{\cp  \in {\mathcal P}_\ocal} \left( - s H_\cp + i \theta_\cp Q_\cp - i n L_\cp\right)}\,.
\ee
We see that $s$ plays the role of a temperature, $i \theta_\cp$ is an imaginary chemical potential and $i n$ is an imaginary `angular velocity'. With (\ref{eq:final}) we have obtained our advertised result (\ref{eq:con2}),
extending the correspondence between Wheeler-DeWitt eigenforms and primon gas partition functions \cite{Hartnoll:2025hly} to cases where the cosmological billiards unfolds in a modular domain of $\H_3$.

We may re-write the partition function (\ref{eq:final}) in a suggestive way as follows.
The partition function is a product of the individual partition functions $z_\cp$ for each complex prime $\cp$. From (\ref{eq:final}), these may be written as
\be\label{eq:CFT}
z_\cp = \tr \left(\left(1/\cp\right)^{A^\cp_{s+n}} \left(1/\bar{\cp}\right)^{A^\cp_{s-n}}\right) \,.
\ee
Here we have set
\begin{align}
A^\cp_x \equiv & \; \half ( x -  i t_\cp) \, b^\dagger_\cp b_\cp + \half(x + i t_\cp) \, c^\dagger_\cp c_\cp \,,
\end{align}
and where
\be\label{eq:tcp}
t_\cp \equiv \frac{\theta_\cp}{\log |\cp|} \,.
\ee
Here we see that (\ref{eq:CFT}) has a certain resemblance to a torus CFT partition function $\propto \text{tr}\left(q^{L_0} \bar q^{\bar L_0} \right)$, with $q$ given by the complex prime $1/\cp$. Furthermore, from our discussion above we have that $s \pm n$ are eigenvalues of $-D \pm i R$, which are proportional to $L_0$ and $\overline{L}_0$ in the $\mathfrak{sl}(2,\mathbb{C})$ algebra. This may suggest that (\ref{eq:CFT}) can be realised as a sector of a modular invariant theory on a torus. The $t_\cp$ coefficients in (\ref{eq:tcp}) appear naturally in the adelic discussion of Appendix \ref{app:adelic}.

\subsection{The simplest complex primon gas}

The expression (\ref{eq:final}) involves terms for every complex prime. We have not used the fact that the primes can be classified as inert, split and ramified. It is not clear to us that this structure will be useful in (\ref{eq:final}) as we do not have enough control over $\arg(\cp)$ and $\theta_\cp$. For completeness we will now briefly describe the simplest complex primon gases, which are the Dedekind zeta functions (see e.g.~\cite{neukirch1999}, ch.7 \S 5). We will see that the structure of the complex primes is useful in this case. For the Gaussian and Eisenstein integers, the Dedekind zeta function is
\be\label{eq:dedekind}
\zeta_\ocal(s) = \sum_{\mu \in \ocal} \frac{1}{|\mu|^s} = n_u \prod_{\cp \in {\mathcal P}_\ocal} \frac{1}{1 - |\cp|^{-s}} \,.
\ee
Recall that $n_u$ is the number of units. Here $|\mu|$ is the usual magnitude of a complex number. We have written (\ref{eq:dedekind}) in a way that brings out the similarities to (\ref{eq:euler2}), although this is not the most conventional normalisation. Like (\ref{eq:euler2}), it is clear that (\ref{eq:dedekind}) is the partition function of a gas of oscillators labeled by complex primes. The oscillators are now neutral.

Consider the ratio of the Dedekind zeta function to the Riemann zeta function. We may use the correspondence between complex primes and ordinary primes, described above, to write
\begin{align}
    \frac{\zeta_\ocal(s)}{\zeta(\frac{s}{2})} &=  n_u \prod_{p \text{ splits}}\frac{1-p^{-\frac{s}{2}}}{(1-p^{-\frac{s}{2}})^2}\prod_{p \text{ inert}}\frac{1-p^{-\frac{s}{2}}}{1-p^{-s}}\prod_{p \text{ ramified}}\frac{1-p^{-\frac{s}{2}}}{1-p^{-\frac{s}{2}}} \label{eq:ff}\\
    &= n_u \prod_{p \text{ splits}}\frac{1}{1-p^{-\frac{s}{2}}}\prod_{p \text{ inert}}\frac{1}{1+p^{-\frac{s}{2}}} \\
    & = n_u \prod_{p \in \P} \frac{1}{1-\chi_\ocal(p)p^{-\frac{s}{2}}} = n_u \sum_{n=1}^{\infty} \frac{\chi_\ocal(n)}{n^\frac{s}{2}} \label{eq:epf} \\
    & = n_u \, L\left(\frac{s}{2}, \chi_\ocal\right) \,.
\end{align}
This is a well-known relation \cite{neukirch1999}.
In these expressions $p$ is an ordinary (`rational') prime.
We defined the different types of prime above and below (\ref{eq:exg}). The contribution from primes that split is squared in $\zeta_\ocal$, these are contributions from a complex prime and its complex conjugate. 
In the final expression $L(s, \chi_\ocal)$ is the Dirichlet $L$-function associated to the Dirichlet characters
\be\label{eq:char}
\chi_{\Z[i]}(n) =
\left\{
\begin{array}{cl}
1 & n = 1 \; (\text{mod } 4) \\
-1 & n = 3 \; (\text{mod } 4) \\
0 & \text{else}
\end{array}
\right.\,,\qquad
\chi_{\Z[\omega]}(n) =
\left\{
\begin{array}{cl}
1 & n = 1 \; (\text{mod } 3) \\
-1 & n = 2 \; (\text{mod } 3) \\
0 & \text{else}
\end{array}
\right. \,.
\ee
These are totally multiplicative functions, which is why the Euler product formula in (\ref{eq:epf}) holds. These two characters are often denoted $\chi_{-4}$ and $\chi_{-3}$, respectively. The results above show that the Dedekind zeta function is a standard primon gas partition function multiplied by an `index' that weights the oscillators according (\ref{eq:char}). See \cite{bakas, Spector1990} for other weighted primon gases.

The Dirichlet characters can be written as
\be
\chi_{\Z[i]}(n) = \sin\frac{n \pi}{2} \,, \qquad \chi_{\Z[\omega]}(n) = \frac{2}{\sqrt{3}} \sin \frac{2 n \pi}{3} \,.
\ee
Using these expressions, the Dirichlet $L$-functions can be written in terms of polylogarithms as
\be
L\left(s, \chi_{\Z[i]}\right) = \text{Im} \, \text{Li}_{s}\left(e^{i \pi/2} \right) \,, \qquad L\left(s, \chi_{\Z[\omega]}\right) = \frac{2}{\sqrt{3}} \text{Im} \, \text{Li}_{s}\left(e^{i 2 \pi/3} \right) \,.  
\ee
These are rather featureless functions at positive $s$, without zeros or poles. Both the Dirichlet $L$-functions and Dedekind zeta function obey a reflection formula \cite{neukirch1999}.

\subsection{Numerical study of Hecke coefficients}
\label{sec:numer}

A large number of eigenvalues and eigenfunctions of the hyperbolic Laplacian in our modular domains have been computed in \cite{Aurich:2004ik} based on Hejhal's algorithm \cite{hejhal1999eigenfunctions}, and improvements thereof \cite{then2005maass}. The arithmetic properties of the domains and the corresponding existence of Hecke operators suggest that the spectral statistics should be Poisson, as is indeed verified numerically \cite{steil1996eigenvalues}.

\begin{figure}[h]
\centering
\includegraphics[width=0.49\textwidth]{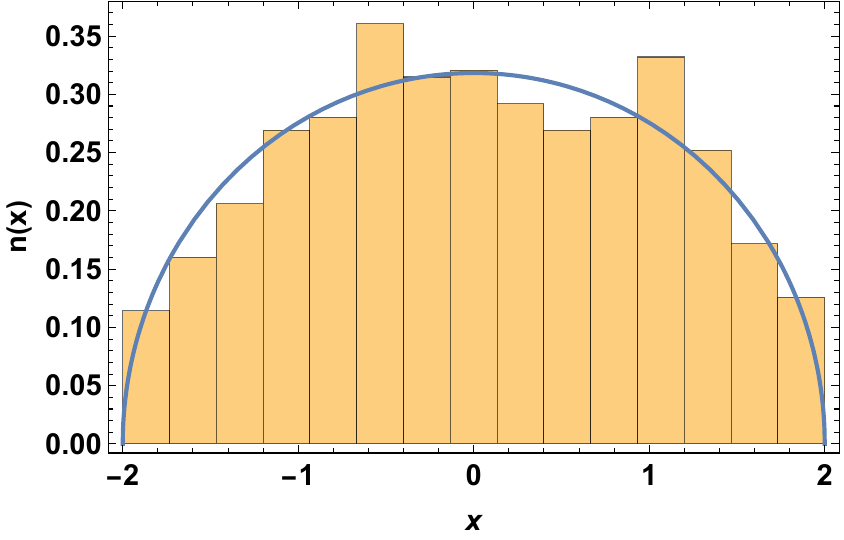}
\includegraphics[width=0.49\textwidth]{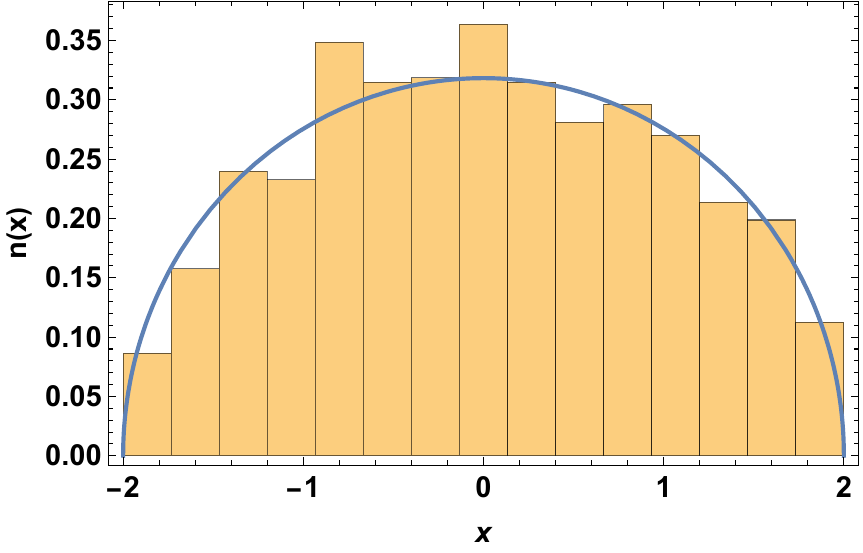}
\caption{Left: Distribution of the first 650 prime Hecke eigenvalues $c_\cp$ associated to the Gaussian waveform at $\varepsilon \approx 25.7239$. Right:
Distribution of the first 1000 $c_\cp$ associated to the Eisenstein waveform at $\varepsilon \approx 24.5033$. In both cases the solid curve shows the Wigner semicircle $n_W(x)$.}
\label{fig: distrbution Hecke eigenvalues}
\end{figure}

While the spectral statistics look integrable, the individual wavefunctions appear random. The in-wavefunction randomness is manifested in the distribution of Hecke eigenvalues $c_\mu$. For a fixed energy level $\vep$ the Sato-Tate conjecture asserts that prime Hecke coefficients should follow a Wigner semicircle distribution. Using Hejhal's algorithm, as described in \cite{then2005maass} and summarised in Appendix \ref{app: hejhal}, we computed prime Hecke eigenvalues associated to a fixed energy eigenfunction for both the Gaussian and Eisenstein billiards. A normalised histogram of these coefficients is shown in Fig.~\ref{fig: distrbution Hecke eigenvalues}. The distribution is indeed consistent with a Wigner semicircle
\be
n_W(x) = \frac{1}{\pi} \sqrt{1 - \frac{x^2}{4}} \,.
\ee

\begin{figure}[h]
\centering
\includegraphics[width=0.49\textwidth]{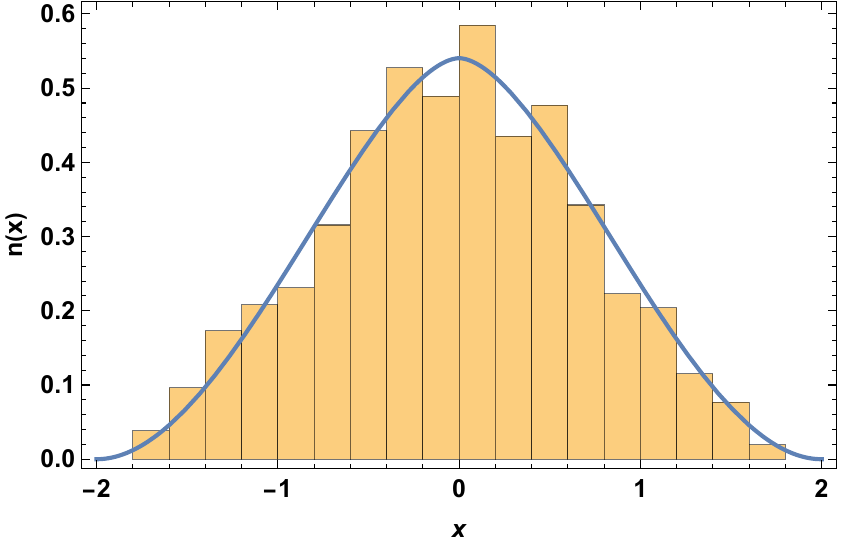}
\includegraphics[width=0.49\textwidth]{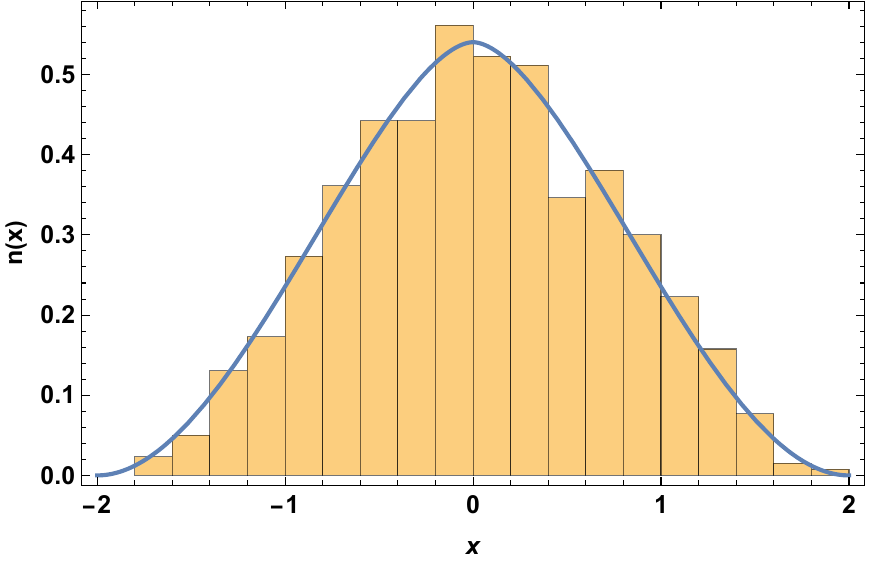}
\caption{Distribution of the first 1300 prime Fourier coefficients $b^-_\cp$ for the two odd Maa{\ss} waveforms
corresponding to Hecke eigenvalues shown in Fig.~\ref{fig: distrbution Hecke eigenvalues}.
The solid curves shows the convolution of two Wigner semicircle distributions.}
\label{fig: distrbution Hecke Fourier}
\end{figure}

The Fourier coefficients $b_\cp^-$ of the Maa{\ss} waveform are obtained as differences of Hecke eigenvalues $c_\cp - c_{\bar \cp}$, according to (\ref{eq: relation an cn}). If $c_\cp$ and $c_{\bar \cp}$ are statistically independent, then $b_\cp^-$ will be distributed as the convolution of two semicircle distributions. Fig.~\ref{fig: distrbution Hecke Fourier} shows that, indeed, the prime Fourier coefficients are distributed according to this convolution.

\subsection{Averaged complex primon gas}
\label{sec:av}

The primon gases (\ref{eq:final}) exist for each energy level $\vep_k$, with different sets of coefficients $\{\theta^k_\cp\}$. Consider a fixed prime $\cp$ and vary the energy level. In this section we restore the label $k$. The coefficients are known to follow the Kesten-McKay distribution
\be\label{eq:KM}
n_\cp(x) = \sum_k \delta(x - c_\cp^k) = \frac{(1 + |\cp|^2) \sqrt{4-x^2}}{2 \pi \left[\frac{1}{|\cp|^2} \left(1+|\cp|^2 \right)^2 - x^2 \right]} \,.
\ee
This is the result in \cite{Sarnak1987}, with $p \to |\cp|^2$ in the complex case as $|\cp|^2$ is the number of equivalence classes modulo the complex prime $\cp$. Recall that $-2 \leq c_\cp^k \leq 2$ by the Ramanujan conjecture. We have verified (\ref{eq:KM}) numerically, see appendix \ref{app: hejhal}.

Following \cite{Hartnoll:2025hly}
we may use (\ref{eq:KM}) to average the logarithm of the primon gas partition functions (\ref{eq:final}) over the energy levels:
\begin{align}
\Big\langle \log \frac{L_n^k(s)}{A_n} \Big\rangle_k 
& = - \sum_{\cp \in \mathcal{P}_\mathcal{O}} \int dx\, n_\cp(x) \log(1-x e^{-in\arg(\cp)}|\cp|^{-s}+e^{-2in\arg(\cp)}|\cp|^{-2s})  \nonumber \\
& =\sum_{\cp \in \mathcal{P}_\mathcal{O}} \frac{|\cp|^2-1}{2}\log \left( 1-e^{-2in\arg(\cp)}|\cp|^{-2(s+1)} \right)\,. \label{eq:Lavg}
\end{align}
Similarly to the case considered in \cite{Hartnoll:2025hly}, the averaged expression (\ref{eq:Lavg}) looks like the logarithm of the Witten index of a fermionic primon gas --- the overall minus sign has disappeared --- with an oscillator degeneracy that grows with the prime magnitude $|\cp|$.

The unaveraged partition function (\ref{eq:Ltot}) is a sum that only converges absolutely for $\text{Re}\, s > 2$. This is the regime in which the Euler product expression (\ref{eq:euler2}) is valid, and indeed the logarithm of (\ref{eq:euler2}) gives a sum over primes that converges absolutely for $\text{Re}\, s>2$. The averaged result (\ref{eq:Lavg}) is obtained in this regime of absolute convergence. However, the averaged logarithm is given in (\ref{eq:Lavg})  by a sum that converges absolutely for $\text{Re}\, s>1$. The improved convergence of the averaged logarithm means that it is possible to produce plots of the averaged quantity down to lower values of $s$ directly using the sum over primes and without analytic continuation. We will do this shortly. However, as noted in \cite{Hartnoll:2025hly} and again below, the extension of the averaged logarithm to lower $s$ can have different analytic properties (divergences) than the analytic continuation of the unaveraged partition functions.

Potential divergences arise from the
large prime contribution to the sum (\ref{eq:Lavg}). These can be extracted using known facts about the asymptotic distribution of rational prime numbers. We may first separate the sum into inert and split rational primes (the single ramified prime is irrelevant for the asymptotic behaviour, and so we ignore it here)
\be
\Big\langle \log \frac{L_n^k(s)}{A_n} \Big\rangle_k 
 =\sum_{p \text{ inert}} \frac{p^2-1}{2}\log \left( 1 - p^{-2(s+1)} \right) + \sum_{p \text{ split}} (p-1)\log \left| 1 - e^{- 2 i n \arg(p)}p^{-(s+1)} \right|\,. \label{eq:sep}
\ee
Here we have used the fact that the inert primes are real and that the split primes produce a complex conjugate pair of complex primes. In a slight abuse of notation we use $\pm \arg(p)$ to denote the argument of the complex primes associated to the split rational prime $p$. We see in (\ref{eq:sep}) that the averaged logarithm is real at real $s$ even while the individual $L$-functions are not real. This occurs because the distribution (\ref{eq:KM}) does not distinguish between a complex prime and its complex conjugate.

It is known that the rational primes are equally divided between the split and inert primes.
This is closely related to Dirichlet's theorem on arithmetic progressions. It follows that both of these sets have prime counting functions $\pi(x) \sim \frac{1}{2} \frac{x}{\log x}$. Furthermore, the Hecke equidistribution theorem says that the angles of split primes are equidistributed on the unit circle as the magnitude of the prime tends to infinity. Thus the asymptotic contribution to (\ref{eq:sep}) is
\begin{align}
\Big\langle \log \frac{L_n^k(s)}{A_n} \Big\rangle_k & \sim - \int^\infty \frac{dx}{\log x} \frac{x^{-2s}}{4} - \int^\infty \frac{dx}{\log x} \frac{x^{-s}}{2} \int_{\frac{-\pi}{n_u}}^{\frac{\pi}{n_u}} \frac{n_u d \theta}{2 \pi} \cos(2 n \theta) \label{eq:av1} \\
& \sim \frac{1}{4} \log \left(s - \frac{1}{2} \right) + \frac{\delta_{n0}}{2} \log \left( s - 1 \right) \,. \label{eq:avL}
\end{align}
The second term in the second line here follows from the fact that, recalling the discussion around (\ref{eq:Avals}) above, the nonzero $n$'s are either multiples of $n_u$ or $\frac{n_u}{2}$. In both cases the final integral in the first line vanishes unless $n=0$. The series with $n \neq 0$ are thus seen to have better convergence properties due to phase cancellations, and can be extended below $s=1$.

\begin{figure}[h]
\centering
\includegraphics[width=0.7\textwidth]{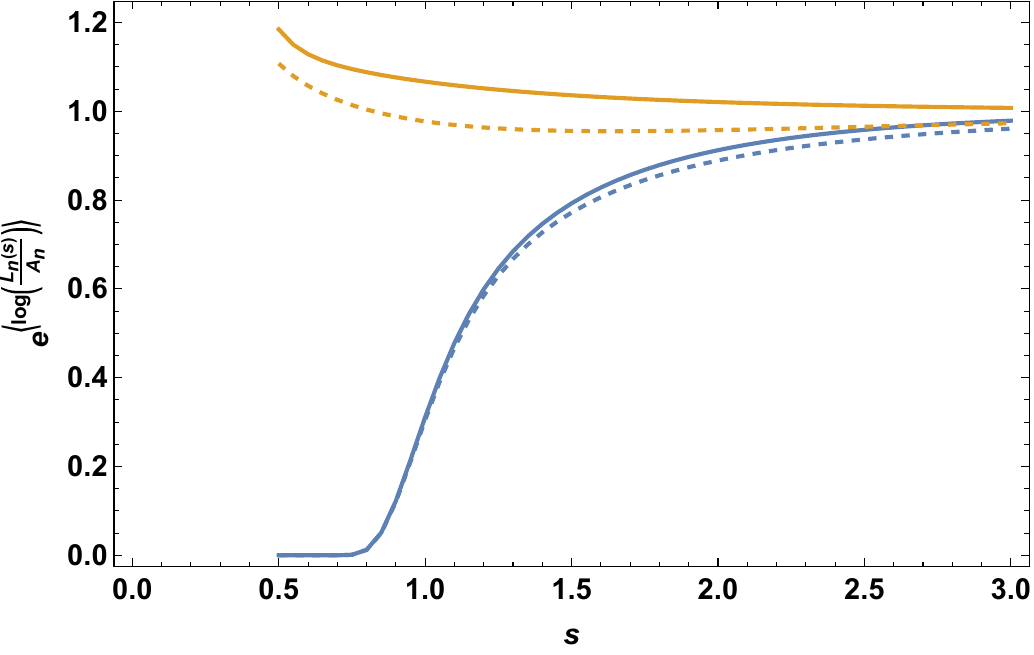}
\caption{Exponential of the averaged primon gas partition functions (\ref{eq:Lavg}). The sums have been performed using all complex primes with $|\cp| < 1000$. Solid curves are from Eisenstein primes and dashed curves are from Gaussian primes. The bottom two curves are for $n=0$ while the top two curves are for $n=3$ (Eisenstein) and $n=4$ (Gaussian).}
\label{fig:avplot}
\end{figure}

The expression (\ref{eq:avL}) suggests that the averaged logarithm of the partition function has divergences at $s=1$ and $s=\frac{1}{2}$. In fact, we will now see that numerical results show that while there is a divergence at $s=1$ when $n=0$, there is no divergence for $n \neq 0$. We believe that this occurs due to subleading corrections to the second term in (\ref{eq:av1}) canceling the divergence in the first term.
One way to get a handle on the subleading corrections is to expand the logarithm in (\ref{eq:Lavg}), to obtain a sum over Hecke $L$-functions --- these have known analytic properties, but one still needs to grapple with the sum. The numerical results are shown in Fig.~\ref{fig:avplot}

In Fig.~\ref{fig:avplot} we plot the exponential of the averaged logarithm. This is a proxy for an averaged $L$-function. Divergences such as (\ref{eq:avL}) in the logarithm lead to zeros in this plot.
The $n=0$ plots in the figure are indeed seen to head towards zero at $s=1$, as suggested by (\ref{eq:avL}). The vanishing is smoothed out due to the truncation of the numerical sum over primes (this sum converges rather slowly). Furthermore, the Eisenstein and Gaussian $n=0$ plots become the same towards $s=1$, consistent with the universal behaviour of the divergence in (\ref{eq:avL}). The plots with $n \neq 0$ in Fig.~\ref{fig:avplot}, in contrast, vanish at neither $s=1$ nor $s=\frac{1}{2}$.

Let us discuss further the vanishing of the $n=0$ curves in Fig.~\ref{fig:avplot}. The unaveraged $L$-functions are not expected to have zeros at $s=1$. This is because (recall \S\ref{sec:hecrelate}) the $L$-functions are built from Hecke eigenforms rather than the Maa{\ss} forms directly. The Hecke eigenforms do not have definite parity and therefore are not forced to vanish anywhere. The exception to this last statement are the $L$-functions with $n = 0$ that are even under $s \leftrightarrow 2-s$, from (\ref{eq:xipm}). The evenness of the $L$-function in this case entails that the $n=0$ Fourier mode of odd Maa{\ss} forms vanishes identically, but does not require the Hecke eigenform $L$-function to vanish at $s=1$. The vanishing of the averaged $n=0$ $L$-function in Fig.~\ref{fig:avplot} is similar to a phenomenon seen in \cite{Hartnoll:2025hly}, where it was noted that even and odd $L$-functions have the same average. That paper also made illustrative plots of individual $L$-functions. We have not been able to make reliable plots here as the individual $L$-functions must be calculated using the full Dirichlet series and these
converge slowly for $s \sim 1$.

\section{Discussion}
\label{sec:final}

As explained in the introduction, this work can be viewed as a step towards obtaining a dual partition function for the quantum cosmological dynamics of ten or eleven dimensional supergravity. One point that we have understood in going from four to five spacetime dimensions is the need to express the WDW wavefunctions in a basis of both dilatation and rotation eigenstates. The next step in this endeavour will be to consider the quaternionic modular symmetry associated to seven dimensional cosmological billiards. A clear challenge here, in relation to prime factorisation, will be the noncommuting nature of quaternions.

Within the $\H_3$ quantum billiards that we have considered in this paper, the suggestive similarity in  (\ref{eq:CFT}) of the fixed prime partition function with a CFT parition function may deserve further consideration.

More broadly, there are several important open questions regarding a WDW approach to cosmological billiards. We will briefly mention a few of these now. Some of these (and others) were discussed in \cite{Hartnoll:2025hly}.

From a gravitational point of view, the decoupling of spatial points deserves greater scrutiny. Away from the strict BKL limit the curvature term in the Hamiltonian constraint equation will couple different points. This term becomes most important close to bounces off of a curvature wall. It should be possible to incorporate these couplings as a perturbation about the decoupled BKL dynamics. It would be very instructive, at both a classical and quantum level, to study the interacting many-body billiards problem that emerges.

A not-entirely-satisfactory aspect of our discussion is that the primon gas partition function is most naturally defined in the regime where the Euler product representation makes sense, $\text{Re}\, s > 2$, while the WDW wavefunction is defined on the critical line $\text{Re}\, s = 1$. In \cite{Hartnoll:2025hly} we obtained an integral transform giving the primon gas partition function directly in terms of the wavefunction. It would be nice to obtain an inverse expression for the wavefunction directly in terms of primon gas quantities, without any need for analytic continuation. This would be analogous to how the bulk wavefunctions of topologically ordered states, such as quantum Hall wavefunctions, are written explicitly in terms of boundary partition function data \cite{RevModPhys.89.025005}. In \cite{Hartnoll:2025hly} we reviewed how the `approximate functional equation' can evaluate the $L$-function on the critical axis in terms of the Dirichlet coefficients. It may be possible to get the primon gas data $\{\theta_\cp\}$ on the critical axis using `symmetrised Euler products' instead of the approximate functional equation, e.g.~\cite{keating1993riemann, gonek2012finite}.

The fact that each WDW eigenfunction potentially connects to two distinct auxiliary quantum systems --- the primon gas and the putative Hamiltonian whose spectrum gives the nontrivial zeros --- appears to be an embarrassment of riches. In \cite{Hartnoll:2025hly} we recalled how the `explicit formula' connects the spectrum of zeros to the prime Dirichlet coefficients, which are the data that defines the primon gas. It may be possible to use the explicit formula to obtain a direct connection between physical quantities in these two systems.

Many considerations of black hole evaporation assume a Schwarzschild interior, in which the Einstein-Rosen bridge grows all the way to the singularity. This has the consequence that the spacelike singularity extends far away from the `corner' that causally connects to the exterior in the final moments of evaporation. See e.g.~Fig.~13 in \cite{Harlow:2014yka} for the causal structure. In contrast, with BKL dynamics the interior volume collapses to zero, while the expanding direction alternates chaotically between the three spatial directions. See \cite{DeClerck:2023fax} for a recent discussion. Therefore, on average, the extent of all three dimensions must individually go to zero. This seems likely to result in all of the singularity being within a Planck length of the `corner' when the black hole becomes Planck sized. This, in turn, may make it easier for the full interior evolution to reconnect back to the exterior. For example, from an exterior point of view the black hole can transition to an excited string state once it evaporates to the string scale \cite{Susskind:1993ws, Horowitz:1996nw, Ceplak:2023afb}. An interior description should be consistent with this phenomenon. Matching the interior and exterior processes will likely be easier if all of the interior ends up accessible to the exterior. This perspective also strongly motivates the inclusion of stringy effects in BKL dynamics.
It would be fascinating to investigate these questions in more detail.

Finally, we have recalled in \S\ref{sec:av} that the prime Fourier coefficients are distributed among different WDW eigenfunctions according to the Kesten-McKay law (\ref{eq:KM}). This distribution encodes combinatorial information on closed walks in large random graphs where each node has degree $|\cp|^2$. This fact may suggest that there is a graph-theoretic interpretation of a typical near-singularity quantum state, possibly leading to an emergent geometrical description. The Sato-Tate semi-circle distribution also suggests, of course, an intriguing link to random matrices and emergent geometry.

\section*{Acknowledgements}

It is a pleasure to acknowledge helpful discussions with Tom Hartman, Jon Keating, Axel Kleinschmidt and Adam Levine. This work has been partially supported by STFC consolidated grant ST/T000694/1. SAH is partially supported by Simons Investigator award \#620869. MY is supported by a Gates Scholarship (\#OPP1144). MDC is supported by a Leverhulme Early Career Fellowship.

\appendix

\section{Adelic perspective}
\label{app:adelic}

In this Appendix we show that the reflection formula~\eqref{eq:xipm} admits an adelic interpretation, as is to be expected for $L$-functions. Our discussion will follow closely that in~\cite{Hartnoll:2025hly} for $SL(2,\Z)$ automorphic primon gases. The additional step is to account for the complex nature of the primes. In an adelic analysis one first defines `local' 
$\Gamma$ factors for every finite place $\mathfrak p$ and for the Archimedean place $\infty$. These factors are Fourier transforms of suitable multiplicative characters, parametrised by $(\theta_{\mathfrak p},n)$ and the complex variable~$s$. The reflection formula is then shown to be equivalent to the `global' statement that the adelic product of these factors equals $1$.

In what follows we briefly review the required completions of $K \equiv \Q[i]$, construct the associated additive characters and use the adelic formalism to re‑express the reflection formula.  For definiteness we focus on Gaussian primes; the Eisenstein case is completely analogous. References for this formalism include \cite{tateref,neukirch1999}.

\paragraph{Archimedean place:} The Archimedean completion of $K$ is $\C$. Given $z \in \C$, the additive character is defined to be
\be
e_{\infty}(z) \equiv \exp\bigl(- 4\pi i\,\text{Re}\, z\bigr) \,.
\ee

\paragraph{Split primes:}  When a rational prime $p$ factors as $p=\mathfrak p\bar{\mathfrak p}$, every $z\in K$ admits the $\mathfrak p$‑adic expansion, for some $\text{ord}_\cp(z) \equiv N$,
\begin{equation}\label{eq:zj}
  z = \sum_{j\ge -N} a_{j}\,\cp^{\,j}, \qquad a_{j}\in\{0,\dotsc,|\cp|^{2}-1\} \,,
\end{equation}
because there are $p = |\cp|^2$ residue classes modulo~$\mathfrak p$. The definition of $N$ is that $a_{-N}$ is the first nonzero coefficient. The $\cp$-adic numbers $K_{\mathfrak p}$ are produced by Cauchy completion with respect to the metric $|z|_\cp=|\cp|^{\ord_\cp(z)}$.  The map
\begin{equation}
  f_{\mathfrak p}: K_{\cp} \longrightarrow \Q_{p}, \qquad
  f_{\cp}\!\Bigl(\sum_{j\ge -N} a_{j}\,\mathfrak p^{\,j}\Bigr) 
               = \sum_{j\ge -N} a_{j}\,p^{j} \,,
\end{equation}
identifies $K_{\mathfrak p}\cong\Q_{p}$, where $\Q_p$ are the ordinary $p$-adics. For $z \in K_\cp$ we therefore set
\begin{equation}\label{eq:ep}
  e_{\mathfrak p}(z) \equiv \exp\bigl(2 \pi i\,\{f_{\mathfrak p}(z)\}_{p}\bigr) \,,
\end{equation}
where $\{\cdot\}_{p}$ denotes the $p$‑adic fractional part. The fractional part is made up of the negative $j$ powers in (\ref{eq:zj}) and defines a real number in the exponent of (\ref{eq:ep}). It is necessary to take the fractional part because the infinite series of positive powers does not define a real number.

\paragraph{Inert primes:}  If $p$ remains inert, write
\begin{equation}
  z = \sum_{j\ge -N} (a_{j}+i b_{j})\,p^{j}, \qquad a_{j},b_{j}\in\{0,\dotsc,p-1\} \,.
\end{equation}
Passing to the Cauchy completion yields $K_{p}\cong\Q_{p}[i]$.  Decomposing $z \in K_p$ into real and imaginary parts defines the additive character
\begin{equation}
  e_{p}(z) \equiv \exp\bigl(4\pi i \{\text{Re} \, z\}_{p}\bigr) \,.
\end{equation}

\paragraph{Ramified prime:} For Gaussian integers the ramified prime is $p = 2 = - i (1+i)^2 = - i\cp^2$. 
The $\cp$-adic expansion and completion work in the same way as the split prime case. That is, for all $z\in K$ we have
\begin{equation}
    z = \sum_{j\ge -N} a_{j}\,\cp^{\,j}, \qquad a_{j}\in\{0,1\} \,.
\end{equation}
The additive character, however, must be defined slightly differently. We again use (\ref{eq:ep}) to define $e_{\cp}(z)$ but now
the map $f_{\mathfrak p}: K_{\cp} \longrightarrow \Q_{p}$ is taken to be
\begin{equation}\label{eq:FFF}
f_{\cp}\!\Bigl(\sum_{j\ge -N} a_{j}\,\mathfrak p^{\,j}\Bigr) 
               = \sum_{j\ge -N} a_{j}\,(\cp^j+\bar{\cp}^j) \,.
\end{equation}
This map is no longer surjective \cite{neukirch1999},
but is still well-defined because $\cp^j+\bar{\cp}^j$ is always a power of 2 and hence the map produces a term in $\Q_2$.

\paragraph{Adelic product:} With the additive characters at hand, we may note that for all $z \in K$,
\begin{equation}\label{eq:ad1}
  \prod_{\a \in \{\infty,\cp\}} e_{\a}(z) = 1 \,,
\end{equation}
The split primes appear in the combination $e_{\cp}(z)\,e_{\bar{\cp}}(z)=\exp(4 \pi i \{\text{Re} \, z\}_p)$, using the fact that $f_{\bar \cp}(z)=f_\cp (\bar z)$. From the construction above, the ramified prime also appears as $\exp(4\pi i\{\text{Re}\  z\}_2)$. Thus (\ref{eq:ad1}) follows from the fundamental (and simple) result in $p$-adic analysis that
\be
\text{Re} \, z - \sum_p \{\text{Re} \, z\}_p \in \Z \,.
\ee

\paragraph{Multiplicative characters:}  For a finite place $\cp$ and $n\in\Z$ let
\begin{equation}\label{eq:multi}
  \chi_{\cp}^{\,n,it}(s,z) \equiv |z|_{\cp}^{\,s+it}\,e^{-in \, \text{ord}_{\cp}(z)\arg(\cp)},
\end{equation}
where again $|z|_{\cp} \equiv |\cp|^{\text{ord}_\cp(z)}$ and ord$_\cp(z)$ was defined above (\ref{eq:zj}), it is the negative of the number of factors of $\cp$ in the prime decomposition of $z$. For the Archimedean place we define
\be
\chi_{\infty}^{\,n,it}(s,z) \equiv |z|^{\,s+it} \left(\frac{z}{|z|}\right)^{-n} \,.
\ee
These multiplicative characters also obey an adelic relation, for all $z \in K$,
\begin{equation}\label{eq:ad2}
  \prod_{\a \in \{\infty,\cp\}}\chi_{\a}^{n,it}(s,z) = 1 \,.
\end{equation}
This follows from considering the prime factorisation of $z$, the same way that the classic adelic formula $\prod_{\a}|x|_{\a}=1$ is established.

\paragraph{Local $\Gamma$ factors:}  For all $\a \in \{\infty,\cp\}$ the local $\Gamma$ factor is the Fourier transform
\begin{equation}
  \Gamma^{n,it}_{\a}(s) 
  \equiv \int_{K_{\a}} \frac{d\mu(z)}{|z|^2_\a}\,\chi^{n,it}_{\a}(s,z)\,e_{\a}(z) \,.
\end{equation}
Here the Archimedean measure is $d\mu(x) = 2 d^2z$ and the measure of the integer $p$-adics $\mu(\Z_\cp) = 1$. Thus we have, for the Archimedean place (cf.~\cite{tateref}, pg.~319),
\be
\Gamma^{n,it}_{\infty}(s) = i^{n} \frac{(2\pi)^{\frac{2-s+|n| - i t}{2}}}{(2\pi)^{\frac{s + |n| + i t}{2}}} \frac{\Gamma\left(\frac{s+|n|+i t}{2} \right)}{\Gamma\left(\frac{2-s+|n|-i t}{2} \right)} \,.
\ee
The $p$-adic integrals may be performed, for non-ramified places (cf.~\cite{tateref}, pg.~322),
\begin{align}
\Gamma^{n,it}_{\cp}(s) & = \sum_{m=0}^\infty \left(\frac{1}{|\cp|^{2m}} - \frac{1}{|\cp|^{2(m+1)}} \right) \frac{1}{|\cp|^{m(s-2 + i t)}} e^{-i n m \arg(\cp)} - |\cp|^{s-2 + i t} e^{i n \arg(\cp)} \\
& = \frac{1 - e^{i n \arg(\cp)} |\cp|^{s-2 + i t}}{1 - e^{-i n \arg(\cp)} |\cp|^{-s - i t}} \,.
\end{align}
In the first line we are summing over the contribution from $z$ with fixed $\text{ord}_\cp(z) = m$. The factor of $\left(1/|\cp|^{2m} - 1/|\cp|^{2(m+1)} \right)$ is the volume of this contribution. The final term in the first line is the contribution with $\text{ord}_\cp(z) = -1$.
The contributions from $\text{ord}_\cp(z) < -1$ vanish.
See e.g.~\cite{Gubser:2016guj, BREKKE19931} for more details on these kinds of integrals. 

For the ramified prime $\cp=1+i$, extra care must be taken in calculating the Fourier transform, which has additional contributions. For the Gaussian integers, there is a contribution from all 
$\text{ord}_\cp(z) \geq -3$.\footnote{
For the ramified prime in the 
Eisenstein integers there is a contribution from all $\text{ord}_\cp(z) \geq -2$.} Furthermore, the failure of the map in \eqref{eq:FFF} to be surjective requires a factor in the measure to compensate for
the fact that $\Q_\cp$ is a bigger set than $\Q_2$. In general, the required factor is $1/\sqrt{|d_K|}$ \cite{neukirch1999, tateref}, here $d_K$ is the discriminant of the field ($d_K = -4$ for Gaussian integers and $d_K = -3$ for Eisenstein integers). The measure must be such that the Fourier transform $F$ obeys $F[F[f]](x)=f(-x)$ \cite{tateref}. This ensures that the transformation is an isometry and that Poisson resummation holds.

We thus have (again, cf.~\cite{tateref}, pg.~322)
\begin{align}
\Gamma^{n,it}_{\mathfrak{p}}(s)
&= \frac{1}{\sqrt{|d_K|}} \left[
  \sum_{m=-2}^{\infty}
    \left( \frac{1}{|\mathfrak{p}|^{2m}} - \frac{1}{|\mathfrak{p}|^{2(m+1)}} \right)
    \frac{1}{|\mathfrak{p}|^{m(s - 2 + i t)}} \,
    e^{-i n m \arg(\mathfrak{p})}
\right. \nonumber \\
&\qquad\left.
  - \frac{1}{|\cp|^2-1}
    \left( \frac{1}{|\mathfrak{p}|^{2m}} - \frac{1}{|\mathfrak{p}|^{2(m+1)}} \right)
    \frac{1}{|\mathfrak{p}|^{m(s - 2 + i t)}} \,
    e^{-i n m \arg(\mathfrak{p})}
    \Bigg|_{m = -3}
\right] \\
& = \frac{1}{2}e^{2in \arg\cp}|\cp|^{2(s+it)}\frac{1 - e^{i n \arg(\cp)} |\cp|^{s-2 + i t}}{1 - e^{-i n \arg(\cp)} |\cp|^{-s - i t}} \,.
\end{align}
In the first line the $m=\{-1,-2\}$  contributions arise because $\cp^{m}+\bar \cp^{m}$ in (\ref{eq:FFF}) has no fractional part in these cases and hence these terms contribute to the integral without cancellations from the additive character (just like the $m \geq 0$ contributions for the non-ramified primes). For $m=-3$, in the second line, the fractional part using (\ref{eq:FFF}) in (\ref{eq:ep}) is
$-1/2$ and hence this term has an overall minus sign (just like the the $m=-1$ contribution for non-ramified primes). In the final line we put $d_K = -4$.

\paragraph{Global reflection formula:} The reflection formula (\ref{eq:xipm}) can be written in the adelic form
\be
\prod_{\a = \{\infty,\cp\}} \Gamma^{n,i t_\a}_{\a}(s) \Gamma^{n,-i t_\a}_{\a}(s) = 1 \,,
\ee
where we set
\be
t_\infty \equiv \vep \,, \qquad t_\cp \equiv \frac{\theta_\cp}{\log|\cp|} \,.
\ee
Note that in (\ref{eq:xi}) the parameter
\be
\a = \frac{4 \pi}{\sqrt{|d_K|}} \,.
\ee
This $\a$ is of course distinct from the place label $\a$ we have been using in this Appendix.

\section{Hecke operators}
\label{app:heck}

In this Appendix we will briefly recall the derivation of the Hecke relations (\ref{eq: hecke relations}). We firstly define the Hecke operators acting on functions on $\mathbb{H}_3$. These operators are labeled by $\nu \in \ocal$,
\be\label{eq:tmu}
T_\nu f(z,y) \equiv \frac{1}{|\nu|} \sideset{}{^\times}\sum_{\substack{a d = \nu \\ b \text{ mod } (d)}} f\left(\frac{a z + b}{d},\frac{|\nu| y}{|d|^2}\right) \,.
\ee
In this sum $a,b,d \in \ocal$ and, as in (\ref{eq: hecke relations}), $d$ is taken modulo units. The sum over $b$ is modulo $(d) = d \cdot \ocal$, the principal ideal generated by the element $d$. The coordinate transformation in (\ref{eq:tmu}) generalises (\ref{eq:expl}), with $c=0$, to allow for elements of $GL(2,\mathcal{O})$ with determinant $\nu$.

The Hecke operators have several important properties. They are self-adjoint on $PSL(2,\mathcal{O})$-invariant functions and hence their eigenvalues are real. In addition, they mutually commute and commute with the Laplacian on $\mathbb{H}_3$. Therefore, there exists a complete set of joint eigenfunctions for the Laplacian and Hecke operators. The Hecke operators are not all independent, but rather satisfy multiplicative relations that we will now establish \cite{steil1996eigenvalues}.

Using the Fourier representation (\ref{eq:cc}) for $f(z,y)$ we obtain
\be
T_\nu f(z,y) = \sum_{\mu \in \ocal} c_\mu \sideset{}{^\times}\sum_{a d = \nu} \frac{y}{|d|^2} \, K_{i\vep}\left(\alpha \left| \frac{\mu \nu}{d^2} \right| y \right) e^{i \alpha \langle \mu,  i \frac{a z}{d}\rangle} \sum_{b \text{ mod } (d)}  e^{i \alpha \langle \frac{i \mu}{d}, b\rangle} \,.
\ee
The final sum over $b$ will add up to zero unless $d$ divides $\mu$, i.e.~unless $\mu/d \in \ocal$. With this condition, $i \a \frac{\mu}{d}$ is in the dual lattice and hence each term in the final sum is $e^{i \alpha \langle \frac{i \mu}{d}, b\rangle} = 1$. This sum is then equal to $|d|^2$, which is the number of nonequivalent Gaussian or Eisenstein integers modulo $(d)$. Thus we obtain
\be
T_\nu f(z,y) = \sideset{}{^\times}\sum_{a d = \nu} \sum_{\frac{\mu}{d} \in \ocal} c_\mu  y \, K_{i\vep}\left(\alpha \left| \frac{\mu \nu}{d^2} \right| y \right) e^{i \alpha \langle \frac{a \mu}{d},  i z\rangle} \,.
\ee
Now define $\mu' = \frac{a \mu}{d}$. It follows that $a$ is a common divisor of $\mu'$ and $\nu$, and we may write
\be\label{eq:Tmusimpl}
T_\nu f(z,y) = \sum_{\mu' \in \ocal} \;\; \sideset{}{^\times}\sum_{a | (\mu',\nu)}  c_{\frac{\mu' \nu}{a^2}}  y \, K_{i\vep}\left(\alpha \left| \mu' \right| y \right) e^{i \alpha \langle \mu',  i z\rangle} \,.
\ee

We now require that $f(z,y)$ be an eigenfunction of $T_\nu$ with eigenvalue $t_\nu$. From (\ref{eq:Tmusimpl}), and recalling the original Fourier expansion (\ref{eq:cc}), this imposes the relation
\begin{equation}
    t_\nu c_\mu = \sideset{}{^\times}\sum_{a|(\mu,\nu)} c_{\frac{\mu\nu}{a^2}} \,.
    \label{eq: hecke eigenvalues}
\end{equation}
Considering the case $\mu = 1$ together with the convention that $c_1=1$ gives the Hecke eigenvalues $t_\nu = c_\nu$. Plugging this result back into (\ref{eq: hecke eigenvalues})
we obtain the Hecke relations
\begin{equation}
    c_\mu c_\nu = \sideset{}{^\times}\sum_{a|(\mu,\nu)} c_{\frac{\mu \nu}{a^2}} \,,
\end{equation}
for any two $\mu, \nu \in \mathcal{O}$. Thus we have obtained (\ref{eq: hecke relations}) in the main text. In the main text we have let $a \to d$, to denote a common divisor.

\section{Hejhal's algorithm}\label{app: hejhal}

\subsection*{Algorithm for the modes}

In this Appendix we review Hejhal's algorithm for the waveforms over $\mathbb{Z}[i]$. The algorithm is similar for $\Z[\omega]$ and is detailed in \cite{Aurich:2004ik}. 
The starting point is the expansion \eqref{eq:Maass} which we now write as
\begin{equation}\label{eq:Maass:app}
    \psi(x_1,x_2,y) = \sum_{\mu \in \mathcal{O^+}} a_\mu y K_{i\vep}( 2\pi |\mu| y) \, f_\mu(x_1, x_2) \,.
\end{equation}
Here $f_\mu$ are given by \eqref{eq:fgau} and the $a_\mu$ obey the symmetries in (\ref{eq:bsym}), so that the wavefunction satisfies the required boundary conditions on the $(x_1,x_2)$ plane. In (\ref{eq:Maass:app}), the symbol $\ocal^+$ means that we sum exclusively over Gaussian integers in the octant $0 \leq \arg(\mu) \leq \frac{\pi}{4}$. Numerically we will have to truncate this sum and consider all $\mu$ with absolute value less than some integer $M$. 

The strategy is to obtain a set of linear equations for the Fourier coefficients $a_\mu$. To do this we will first invert (\ref{eq:Maass:app}) using a Discrete Fourier Transform (DFT) and then use the modular invariance of $\psi$. The DFT gives
\begin{equation}\label{eq:DiscreteFourierTransform}
    a_\mu y K_{i\vep}( 2\pi |\mu| y)  = \frac{2}{Q^2} \sum_{l,k=1}^Q \psi(\tilde{x}_l,\tilde{x}_k,y) f_\mu(\tilde{x}_l,\tilde{x}_k) \,, \qquad \tilde{x}_l = \frac{1}{2 Q} \left(l-\half \right) \,.
\end{equation}
Given (\ref{eq:Maass:app}) with a cutoff $M$ on $|\mu|$, this is an exact expression for any $Q > M$.

We have already imposed the symmetries (\ref{eq:bsym}) on the Fourier coefficients. The full modular invariance of $\psi$ furthermore requires that (\ref{eq:DiscreteFourierTransform}) remain valid if we substitute $\psi(z,y) \to \psi(\gamma(z,y))$ for any modular transformation $\gamma \in PSL(2,\mathbb{Z}[i])$, including inversions. We may then Fourier expand the transformed wavefunction $\psi(\gamma(z,y))$ using (\ref{eq:Maass:app}). 
Furthermore, we may do a different transformation $\gamma_{lk}$ for every point $(\tilde x_l, \tilde x_k)$ in the sum. Thus we set $(x_l^*,x^*_k,y_{lk}^*) \equiv \gamma_{lk}(\tilde x_l,\tilde x_k,y)$ and obtain from (\ref{eq:DiscreteFourierTransform}) that
\begin{equation}\label{eq: linear equation}
    a_\mu y K_{i\vep}( 2\pi |\mu| y)  = \frac{2}{Q^2} \sum_{l,k=1}^Q \sum^M_{\nu \in \mathcal{O}^+} a_\nu y_{lk}^* K_{i\vep}( 2\pi |\nu| y_{lk}^*) f_\nu(x^*_l, x^*_k) f_\mu(\tilde{x}_l, \tilde{x}_k) \,.
\end{equation}
This is a linear system of equations for the Fourier coefficients $a_\mu$, and it should hold for all $y$ whenever $\vep$ corresponds to an eigenvalue of the Laplacian.

We are free to scale the overall waveform, and we use this freedom to set $a_{2+i} = 1$, recalling that $a_1 = 0$ from the symmetries (\ref{eq:bsym}). The linear system of equations (\ref{eq: linear equation}) is then uniquely solved for the remaining Fourier coefficients.

To run the algorithm, we must choose values of $y$ and $M$. Firstly, recall that the modified Bessel function decays exponentially at large arguments, 
$K_{i \vep}(x) \sim x^{\frac{1}{2}} \exp(-x)$. Furthermore, recall that points in the fundamental domain have $y \geq y_0 = \frac{1}{\sqrt{2}}$. These two facts motivate the following. Start with some initial $y_i < y_0$ and find, for each point $(\tilde z,y_i)$ in \eqref{eq:DiscreteFourierTransform}, the unique $PSL(2,\mathbb{Z}[i])$ transformation that maps this point into the fundamental domain. Under this transformation, the $y$ coordinate of all of the images will be larger than the initial $y_i$. From the decrease of the Bessel function it follows that the error due to truncating the sum at $M$ can only be improved through the use of these modular transformations. Note that the required transformations can easily be found by requiring the transformed $y$ coordinate to be maximal, using \eqref{eq:expl}. In conclusion, we take $y_i < y_0$. 

Finally, we require $M y_0$ to be large enough that the truncated sums are a good approximation, but not so large that the matrix appearing in the linear problem becomes ill-conditioned. We may recall that the maximum of $K_{i \vep}(x)$ occurs at $x \sim \vep$ and has magnitude $\sim e^{-\frac{\vep \pi}{2}}$. We therefore pick a small number $\epsilon \ll 1$ and choose $M$ and $y_i$ such that
\begin{equation}
    |K_{i \vep}(2 \pi M y_0)| e^{\frac{\vep\pi}{2}} \approx \epsilon \qquad \text{and} \qquad y_i = \frac{\vep}{2\pi M} \,.
\end{equation}
If the solution to the linear system of equations is independent of the choice of $y_i$, to a prescribed precision, then $\vep $ is an eigenvalue of the Laplacian and the algorithm provides the associated Fourier coefficients $a_\mu$.
We may note that this algorithm determines all of the coefficients independently. After obtaining the coefficients $c_\mu$ of the Hecke eigenforms from the mode coefficients $a_\mu$, as described below, we may then further check the accuracy of the results by verifying the Hecke relations (\ref{eq: hecke relations}) between the $c_\mu$.

\subsection*{Relating the modes and Hecke eigenforms}

The previous section described how to compute, numerically, the Fourier coefficients of a reflection-odd wavefunction. Recall that we focused on the case of $PSL(2,\Z[i])$ and set $a_{2+i} = 1$. In this section we describe how to obtain the corresponding Hecke eigenvalues $c_\mu$. This amounts to an inversion of the relation (\ref{eq: relation an cn}). It is convenient to relax this relation slightly, to allow for a uniform rescaling of the coefficients. Thus we are looking for $c_\mu$ such that
\be\label{eq:alpha}
2 A \, a_\mu = c_\mu - c_{\bar \mu} \,,
\ee
for some constant $A$ such that $c_1 = 1$ and $a_{2+i}=1$.

Using (\ref{eq:alpha}) repeatedly together with the Hecke relations \eqref{eq: hecke relations} we may write
\begin{align}
2 A \, a_{(2+i) \mu} = c_{(2+i)\mu} - c_{(2-i) \bar \mu}
 & = c_{2+i} c_\mu - c_{2-i} c_{\bar \mu} - c_{\mu/(2+i)} + c_{\bar \mu/(2-i)} \\
 & = 2 A \, \left( c_{2+i} a_\mu + c_{\bar \mu} - a_{\mu/(2+i)} \right) \,. \label{eq:e1}
\end{align}
The terms such as $c_{\mu/(2 + i)}$ are defined to be zero if $2+i$ does not divide $\mu$. Using the fact that $a_{\bar \mu} = - a_\mu$, we also have that
\be
a_{(2+i) \bar \mu} = - c_{2+i} a_\mu + c_{\mu} - a_{\bar \mu/(2+i)} \,.\label{eq:e2}
\ee
Adding (\ref{eq:e1}) and (\ref{eq:e2}) gives
\begin{equation}
    c_\mu+ c_{\bar \mu}= a_{(2+i)\mu} +  a_{(2+i)\bar{\mu}} + a_{\mu/(2+i)}+ a_{\bar{\mu}/(2+i)} \,.
\end{equation}
Then using (\ref{eq:alpha}) one last time to eliminate $c_{\bar \mu}$ we obtain
\begin{equation}
   c_\mu= \half \left( a_{(2+i)\mu} +  a_{(2+i)\bar{\mu}} + a_{\mu/(2+i)}+ a_{\bar{\mu}/(2+i)} \right) + A \, a_{\mu} 
   \label{eq: Hecke eigenvalues from data}
\end{equation}
which holds for any integer $\mu$. With (\ref{eq: Hecke eigenvalues from data}) we have achieved our objective of a formula for $c_\mu$ in terms of the $a_\mu$.

The constant $A$ can now be determined using any of the Hecke relations. For example, 
\begin{align}
    a_{10+5i} - a_{2+i} = c_5 = c_{2+i}c_{2-i}= \left(\half a_{(2+i)^2} + A \, a_{2+i} \right) \left(\half a_{(2+i)^2} - A \, a_{2+i} \right)
\end{align}
where we used \eqref{eq: Hecke eigenvalues from data} twice and used $a_{\bar \mu} = - a_\mu$. This yields a quadratic equation for $A$. Recalling that $a_{2+i} = 1$ we obtain
\begin{equation}
    A = \pm \sqrt{1- a_{10+5i}+{\textstyle \frac{1}{4}}a^2_{(2+i)^2} }\,.
\end{equation}

\subsection*{Verifying the Kesten-McKay distribution}

\begin{figure}[h]
\centering
\includegraphics[width=0.49\textwidth]{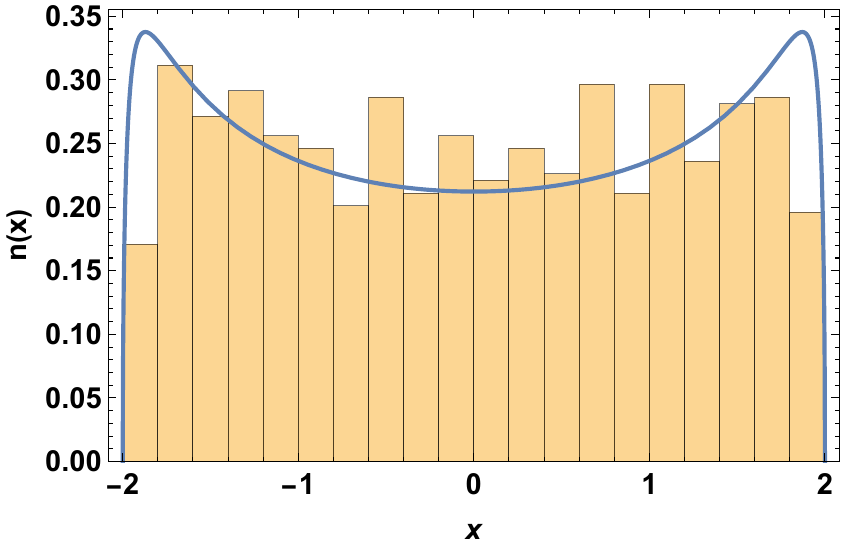}
\includegraphics[width=0.49\textwidth]{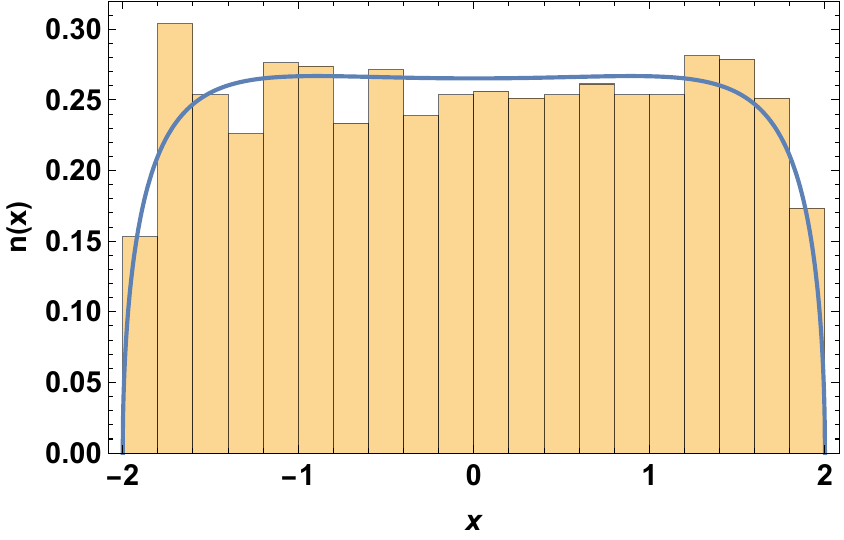}
\includegraphics[width=0.49\textwidth]{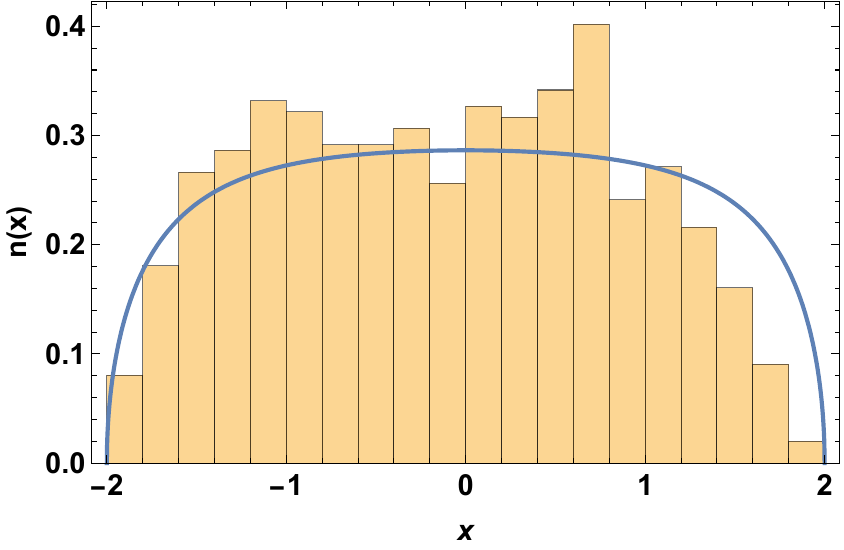}
\includegraphics[width=0.49\textwidth]{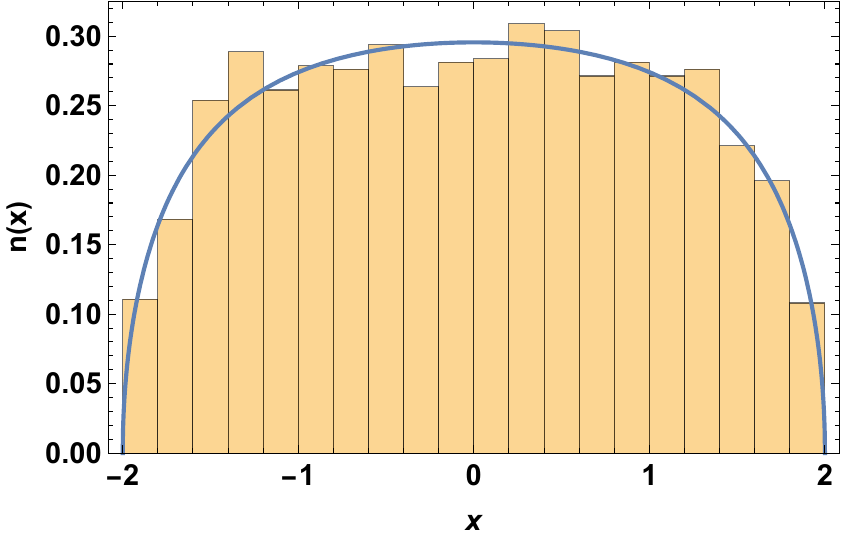}
\caption{Distribution of fixed $\cp$ Hecke eigenvalues $c^k_\cp$ across 995 different energy levels for the Gaussian billiard. Starting at the top left the primes plotted are $\cp = 1+i, 2 \pm i, 3, 3 \pm 2 i$. The solid curve is the corresponding Kesten-McKay distribution (\ref{eq:KM}). As noted in the text, the split primes must be considered together with their complex conjugate.}
\label{fig:verify}
\end{figure}

The plots in Fig.~\ref{fig:verify} are a check on the Kesten-McKay distribution (\ref{eq:KM}), and also on our numerics. It is numerically intensive to make these plots because the Hecke eigenvalues must be found for many different energy levels. While the histograms show significant fluctuations, they are consistent with the magnitude and trend of the distribution as $|\cp|$ is varied. The fact that the Kesten-McKay distribution only depends on the magnitude $|\cp|$ means that split primes and their complex conjugates, $\cp$ and $\overline{\cp}$, must be combined into the same plot.

\providecommand{\href}[2]{#2}\begingroup\raggedright\endgroup

\end{document}